\documentclass[reprint, NumberedRefs]{JASA}
\usepackage{multirow}
\makeatletter
\def\ps@titlepage{%
  \def\@oddhead{\hfill}\let\@evenhead\@oddhead
  \def\@evenfoot{{\footnotesize\sf
    \ifodd\c@page\hfill\thepage\else\thepage\hfill\fi}}%
  \let\@oddfoot\@evenfoot
}
\def\titleblock@produce{%
 \begingroup
  \ltx@footnote@pop
  \def\@mpfn{mpfootnote}%
  \def\thempfn{\thempfootnote}%
  \c@mpfootnote\z@
  \let\@makefnmark\frontmatter@makefnmark
  \frontmatter@setup
  \thispagestyle{titlepage}\label{FirstPage}%
  \frontmatter@title@produce
  \groupauthors@sw{\frontmatter@author@produce@group}{\frontmatter@author@produce@script}%
  \frontmatter@RRAPformat{%
   \expandafter\produce@RRAP\expandafter{\@date}%
   \expandafter\produce@RRAP\expandafter{\@received}%
   \expandafter\produce@RRAP\expandafter{\@revised}%
   \expandafter\produce@RRAP\expandafter{\@accepted}%
   \expandafter\produce@RRAP\expandafter{\@published}%
  }%
  \frontmatter@abstract@produce
  \vskip-4pt
  \@ifx@empty\@pacs{}{\@pacs@produce\@pacs}%
  \@ifx@empty\@keywords{}{\@keywords@produce\@keywords}%
  \par
  \frontmatter@finalspace
 \endgroup
 \pagestyle{reprint}%
}
\makeatother

\begin{document}

\title[]{Localized time-frequency representation learning for bioacoustic classification in complex soundscapes}

\author{Simen Hexeberg}
\affiliation{ARL, Tropical Marine Science Institute, National University of Singapore}
\affiliation{Department of Electrical \& Computer Engineering, National University of Singapore}

\author{Mandar Chitre}
\email{mandar@nus.edu.sg}
\affiliation{ARL, Tropical Marine Science Institute, National University of Singapore}
\affiliation{Department of Electrical \& Computer Engineering, National University of Singapore}

\author{Matthias Hoffmann-Kuhnt}
\affiliation{ARL, Tropical Marine Science Institute, National University of Singapore}

\author{Bing Wen Low}
\affiliation{National Parks Board, Singapore}

\begin{abstract}

Prevailing bioacoustic classifiers assign species labels to fixed time-frequency windows rather than to individual vocalizations. When multiple vocalizations occur within the same window, predictions cannot be unambiguously linked to specific calls, which limits analyses at the level of individual vocalizations. This work introduces a framework for time-frequency localized bird classification. A Local-Context Classifier (LCC) identifies species from localized time-frequency events (TFEs), while a Dual-Context Classifier (DCC) combines local and global acoustic context through a fine-tuned bioacoustic foundation model. On an in-distribution dataset from Singapore comprising 306 vocalization classes from 103 bird species, the LCC achieves an F1-micro score of 79.3\%, while combining local and global context through the DCC yields the highest overall performance (94.6\%). To reduce labeled data requirements, the LCC is pre-trained via self-supervised contrastive learning, achieving an 18.8\% relative gain on an out-of-distribution dataset. A focused evaluation on continuous soundscape recordings further demonstrates the potential of the framework for long-term monitoring applications. By preserving the time-frequency localization of individual vocalizations, the proposed framework supports both ecological monitoring and vocalization-level studies of animal acoustic behavior.
\end{abstract}

\maketitle

\section{Introduction}
\label{sec:introduction}
Biodiversity monitoring is a critical aspect of biodiversity conservation, as it helps inform decision making, improves our knowledge and enhances public education and awareness. Birds are one of the most surveyed animal groups in biodiversity monitoring programs, with point counts and transect surveys being well-established survey techniques for monitoring bird communities~\cite{bibby2000}. However, birds can be very difficult to detect and identify especially in tropical regions characterized by dense foliage, high avian diversity and numerous rare species~\cite{Robinson2000, Robinson2018}. Additionally, such manned survey techniques are manpower-intensive, require highly specialized expertise, and tend to overlook rare species that are sensitive to human presence~\cite{darras2018, Darras2019, Wheeldon2019}.

Passive monitoring of biodiversity using acoustics is thus an area of great potential, as various animal groups including birds make unique vocalizations, which can be used to validate their presence. Such systems allow for automated collection of large amounts of audio data without human supervision and can survey cryptic species more effectively~\cite{darras2018, Hoefer2023}. However, the comprehensive analysis of such large volumes of data is prohibitive in terms of the man-hours required~\cite{Hingston2018}. This constraint, and the rapid advancement in machine learning techniques, have made data-driven algorithms increasingly popular for bioacoustic species detection and classification tasks.

Besides serving as indicators of ecosystem health, birds are also of interest from a machine learning perspective, as bird vocalizations have been shown to be a good source for bioacoustic models to generalize across taxa, extending beyond birds to groups such as amphibians and marine mammals~\cite{Ghani_2023}. However, automatically monitoring birds can be especially challenging during chorus periods, when many individuals vocalize simultaneously. This challenge is amplified in highly biodiverse regions such as tropical ecosystems, where ecological monitoring is particularly important~\cite{Martínez2025}. These dense and overlapping soundscapes are precisely where current bioacoustic classifiers face limitations. Many approaches are single-label methods~\cite{Gupta2021, Thakur2019, Nolasco2023, Manriquez2024}, meaning they can only identify one species within a given time window. Multi-label classifiers~\cite{KAHL2021, LeBien2020, Noumida2022} address this shortcoming by allowing multiple species to be recognized simultaneously. However, both single- and multi-label methods commonly rely on \textit{global-context} representations. Although earlier work has explored localized time-frequency representations~\cite{Brandes2008, Briggs2012}, recent state-of-the-art bioacoustic classifiers typically operate on global-context inputs, most commonly Mel spectrograms~\cite{schwinger2025, Stowell2022}. Such classifiers can detect multiple species within a recording but cannot precisely localize detections in time and frequency or associate predictions with specific vocalizations when multiple vocalizations occur within the same input window. Moreover, the input windows used by these models are typically fixed and model-specific~\cite{schwinger2025}, despite vocalizations naturally varying in both temporal and spectral extent.

To address these limitations, we propose a Local-Context Classifier (LCC) that learns directly from localized time-frequency representations rather than broadband fixed-window inputs. Because local context alone may omit broader temporal cues relevant for species classification, we further propose a Dual-Context Classifier (DCC), which combines LCC with a fine-tuned version of BirdNET~\cite{KAHL2021}---a state-of-the-art bioacoustic foundation model operating on global context. The resulting framework enables vocalization-level detections and supports bioacoustic studies in complex soundscapes where precise time-frequency localization of sound sources is important---for example, studies of how animals adapt their vocalizations in response to changes in their acoustic environment~\cite{both2012, medeiros2017, hopkins2023, farina2012}, including introduced anthropogenic noise. The framework may also support population estimation for species with individually distinctive call signatures, providing an alternative or complement to triangulation-based approaches~\cite{Hexeberg2023}.

Beyond source isolation, the LCC offers two related but distinct advantages. First, it enables multi-label inference without requiring training on multi-label data. This is advantageous for two reasons: (i) single-label classification is generally a simpler learning objective and may require less training data, and (ii) multi-label datasets are costly to annotate~\cite{Briggs2012} and remain relatively scarce in the open-source community---a limitation that motivated the release of the BirdSet dataset~\cite{BirdSet}. Second, it facilitates efficient retrieval of similar vocalizations from long-duration continuous recordings. Together with time-frequency localization, this can reduce annotation effort when constructing multi-label datasets.

To date, most passive acoustic monitoring studies have focused on pristine habitats, while largely overlooking urban ecosystems~\cite{Fairbrass2017, Gibb2019}. As urban green spaces become increasingly important for bird populations under rapid urbanization~\cite{Hughes2022}, evaluating the efficacy of automated analysis in urban soundscapes is particularly urgent. Singapore---one of the few tropical cities with a connected network of green spaces adjacent to dense urban areas~\cite{Wong2023}---is therefore an ideal study site. The proposed models are trained on soundscapes from Singapore and evaluated both in-distribution within Singapore and out-of-distribution on recordings collected outside Singapore. Finally, to assess deployment potential in a realistic long-term monitoring scenario, a focused evaluation is conducted on continuous soundscape recordings from two sites within the Singapore Botanic Gardens (SBG), one of the oldest botanic gardens in Southeast Asia, receiving millions of visitors annually.

\section{Methodology}
This section describes the methods and experimental setup used in this study. We begin by describing the study site and recording setup in Section~\ref{sec:site}. We then introduce the Local-Context Classifier, a three-step pipeline that segments and classifies individual bird calls from acoustic recordings (Section~\ref{sec:lcc}). Next, we present fine-tuned versions of BirdNET (Section~\ref{sec:ftbn}) before BirdNET and the LCC are combined in a Dual-Context Classifier (Section~\ref{sec:dcc}). Finally, we describe the datasets used for training and evaluation (Section~\ref{sec:datasets}), and the metrics used to assess performance (Section~\ref{sec:metrics}).

\subsection{Study site and recording setup}
\label{sec:site}

\begin{figure*}[t]
\centerline{\includegraphics[width=\linewidth,keepaspectratio]{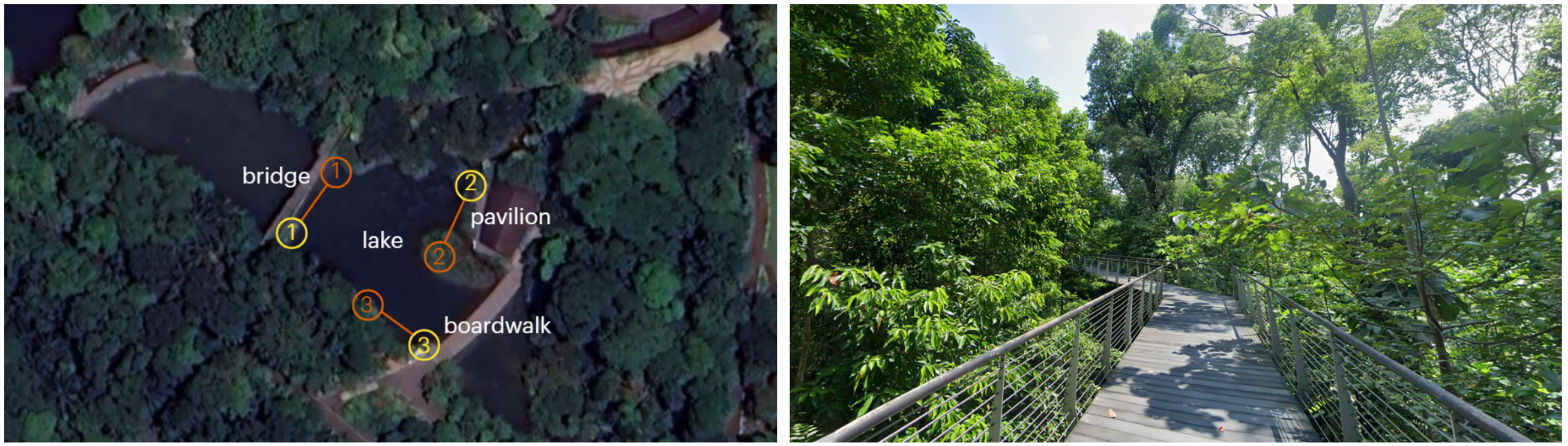}}
    \caption{Left: first deployment at the SBG (site~\#1) from July 4, 2020 until September 20, 2020. The approximate locations of the recording units are marked as yellow circles, with the corresponding external microphones of each unit marked as orange circles. The six microphones surround a lake and cover an area of roughly 50~m $\times$ 50~m. Right: a section of the elevated boardwalk used for the second deployment at the SBG (site~\#2) from September 20, 2020 until February 1, 2021. The same six microphones from site~\#1 were deployed along the circular boardwalk in a similar constellation, but without direct line of sight between units due to the dense vegetation. Photos from Google Maps.}
    \label{fig:sbg_sites}
\end{figure*}

Our primary source of acoustic data was acquired from two different locations in the SBG over a combined period of 7~months from July 2020 to February 2021. Three \emph{Wildlife Acoustics SongMeter 4 TS} recorders were deployed at both locations. Each recording unit was equipped with two omni-directional microphones, yielding soundscape recordings from six microphones simultaneously. The first 2.5~months of data collection took place around a lake with minimal obstruction between the microphones (site~\#1). This setup allows the same call to be detected on multiple recorders (Figure~\ref{fig:sbg_sites}), which we later exploit during model training. The units were then relocated to a second site for the remaining 4.5~months (site~\#2). A similar constellation was used but dense vegetation occluded the direct paths between microphones (Figure~\ref{fig:sbg_sites}). During calibration tests at site~\#2, we emitted high-frequency transients and found that the sounds were mostly audible only on the nearest few microphones. Although low-frequency sounds may attenuate less, site~\#2 is likely to have a shorter detection range than site~\#1.

\subsection{Local-Context Classifier}
\label{sec:lcc}
 The Local-Context Classifier is a three-step method:
 
\begin{enumerate}
\item \textbf{Time-frequency events:} Extract individual bird calls using an energy-based segmentation technique. Isolating individual calls limits noise and allows time-overlapping vocalizations to be treated separately, provided they do not also overlap in frequency.\label{item:segmentation}
\item \textbf{Embedding representation:} Learn compact embedding representations of the time-frequency segments from step~\ref{item:segmentation}, encouraging embeddings that are translationally invariant and that place acoustically similar sounds close together---two key properties for efficient clustering and classification of bioacoustic sounds.\label{item:clr}
\item \textbf{Supervised classification:} Feed the learned embeddings to a classification head and train it on labeled data. This step also serves as a final refinement of the embedding space.
\end{enumerate}

Note that only the last step requires labeled data: step~\ref{item:segmentation} does not involve any learning while step~\ref{item:clr} is self-supervised. Each step is detailed in the next sections.

\subsubsection{Time-frequency events}
\label{sec:tfe}

\begin{figure*}[t]
\centerline{\includegraphics[width=\linewidth,keepaspectratio]{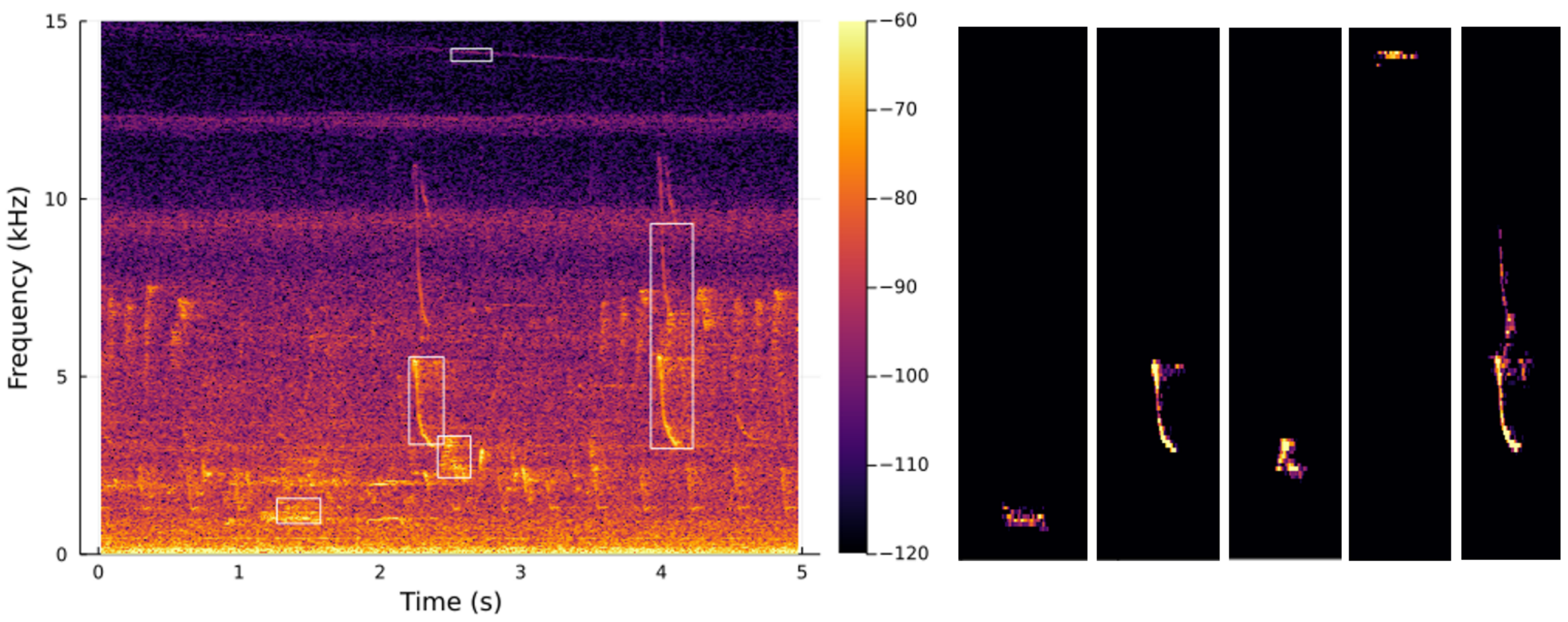}}
    \caption{An example illustrating extraction of TFEs from audio recordings. The left panel shows the spectrogram from a 5-second audio clip, with a few detected sounds enclosed by white rectangles. The right panels show the respective extracted TFEs. This example is non-exhaustive, i.e., not all detections in the audio clip are shown here.}
    \label{figTfrSplit}
\end{figure*}

Most bird vocalizations are frequency-modulated signals. While vocalizations may overlap in time, they may be separated in frequency. Rather than training on broadband, fixed-duration spectrograms---which in dense soundscapes often contain vocalizations from multiple species---we extract individual acoustic events and process them separately. Related acoustic-focused approaches have previously localized animal vocalizations in time and frequency before classification. Neal et al.~\cite{neal2011} and Briggs et al.~\cite{Briggs2012} used supervised time-frequency segmentation to separate bird sounds from background noise in spectrograms. More recently, object-detection approaches have been used to localize vocalizations with bounding boxes rather than pixel-level masks~\cite{Hexeberg2023}. While this reduces annotation effort, it still requires manually annotated training data. Ruiz-Mu\~noz et al.~\cite{ruiz2015} avoided the annotation requirement by using unsupervised spectrogram segmentation to generate candidate regions for birdsong classification. However, that method relies on handcrafted features extracted from the segmented regions and performs classification at the recording level. In contrast, we use an annotation-free, energy-based time-frequency segmentation technique followed by self-supervised feature learning, enabling classification at the level of individual extracted events. The resulting segments are represented as $128 \times 256$ numerical matrices, which we refer to as time-frequency events (TFEs). TFEs are extracted from acoustic recordings through the following steps:

\begin{enumerate}
    \item Compute a spectrogram of the acoustic data with 2,048 FFT bins, a Hamming window, and an overlap of 1,536 samples between windows. Retain frequencies between 500~Hz and 15~kHz, as this range encompasses most bird sounds while rejecting unwanted noise.
    \item Convert the spectrogram to dB and use the inter-quartile range for each frequency bin to obtain a robust estimate of the noise variance $\sigma$ at that frequency. Normalize each frequency bin by subtracting the median + $2\sigma$, dividing by $2\sigma$, and clip the resulting data between 0 and 1. This adaptively extracts regions in the time-frequency plane that have significantly higher energy than the background noise at that frequency. Moreover, it reduces the natural dominance of low-frequency calls resulting from higher attenuation of high-frequency signals~\cite{Sutherland1998}.
    \item Reduce the frequency resolution by a factor of 5 by max-pooling.
    \item Zero out time bins with low variance across broad frequency bands, as these represent impulsive sounds not characteristic of birds.
    \item Perform a watershed segmentation of the resulting spectrogram to obtain disconnected regions of high energy in the spectrogram.\label{item:watershed}
    \item Filter out regions with very short durations or very small time-bandwidth products, as these are uncharacteristic of bird sounds.    
\end{enumerate}

The end result of the above steps is illustrated by an example in Figure~\ref{figTfrSplit}. The spectrogram on the left shows a 5-second recording with multiple bird vocalizations. After segmentation, the spectrogram is converted to a number of TFEs. A few of these TFEs are shown in the panels on the right with corresponding sections marked with white rectangles in the original spectrogram on the left. Although the heuristic rules filter out many events uncharacteristic of birds, the remaining TFEs are not guaranteed to contain only bird sounds. For example, frequency-modulated, high-energy sounds such as insect noise and human speech may pass through these filters and must be handled downstream. Furthermore, TFEs may not capture complete calls when successive syllables within a vocalization are not sufficiently contiguous in time and frequency to be grouped into a single segment.

Although TFEs have variable durations, several algorithms in our processing chain require fixed-length input. We therefore standardize all TFEs to a duration of 2.7~seconds (256 time bins) as most TFEs fall within this range. The few TFEs longer than 2.7~seconds are cropped to 256 bins, while shorter ones are randomly zero-padded on both sides. The random padding and cropping is applied at runtime to create different versions of the TFE each time it is used during training.

\subsubsection{Embedding representation}
\label{sec:clr}

Although TFEs provide a localized representation of individual acoustic events, they are not necessarily optimal input representations for classification or clustering. First, TFEs are sparse: most entries have zero energy. Feeding sparse representations directly to a fully connected neural network would be computationally inefficient, as most inputs would contain no signal. Second, bird vocalization classification is time-invariant: shifting a vocalization in time does not alter its class. The objective of this stage is therefore to learn compact \textit{embeddings} of TFEs that are invariant to time translations and map acoustically similar sounds to nearby points in the embedding space.

Several approaches have been proposed for learning bioacoustic embeddings. One strategy is to use the internal representations of large pre-trained models by removing the classifier head, an approach that has been shown to support transfer learning for bioacoustic classification~\cite{Ghani_2023}. Contrastive learning has also been explored for bioacoustic tasks, including few-shot bird-sound classification~\cite{Moummad2024} and bioacoustic event detection~\cite{Bermant2022}. Our setting differs in that we seek not only self-supervised representations, but also embeddings that preserve time-frequency localization and support both classification and retrieval of similar vocalizations in continuous soundscapes.

\begin{figure*}[t]
  \includegraphics[width=\linewidth]{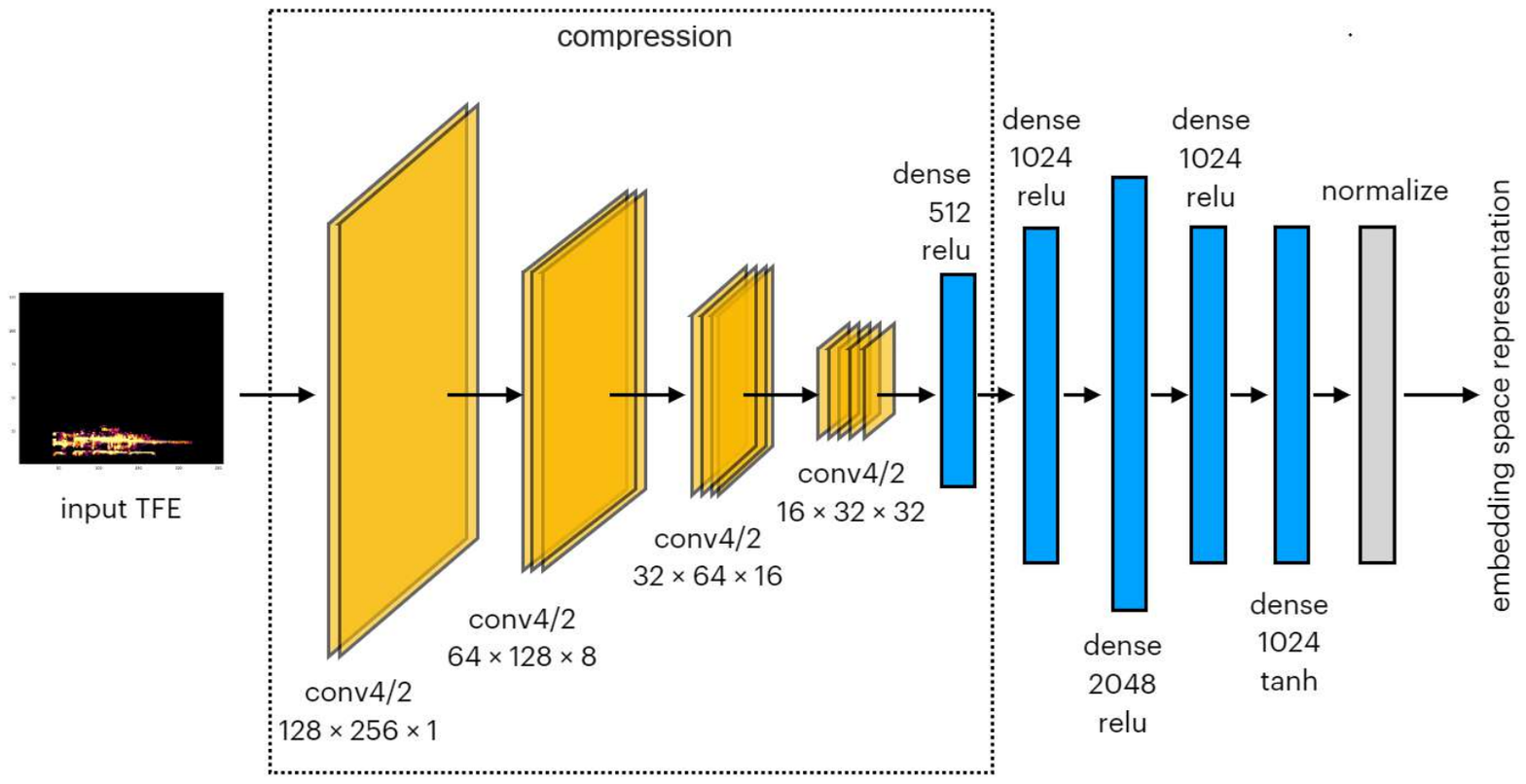}
  \caption{Architecture of the self-supervised contrastive neural network. Features are extracted from the input TFE, compressed, and mapped to a 1024-dimensional embedding space. The embeddings are normalized to lie on the surface of a unit hypersphere, allowing similarity between embeddings to be measured by their dot product, or equivalently by the angle between them.}
  \label{fig:clr}
\end{figure*}

To achieve this, we employ a two-stage neural network architecture (Figure~\ref{fig:clr}). The first stage comprises a convolutional neural network that extracts local time-frequency features from TFEs, followed by a dense layer that projects these features into a lower-dimensional latent space. Specifically, each $128 \times 256$ TFE is compressed into a 512-dimensional latent vector, reducing the representation size by a factor of 64. The second stage consists of four fully connected layers that transform the latent representation produced by the first stage into the final embedding.

\begin{figure*}[t]
  \includegraphics[width=\linewidth]{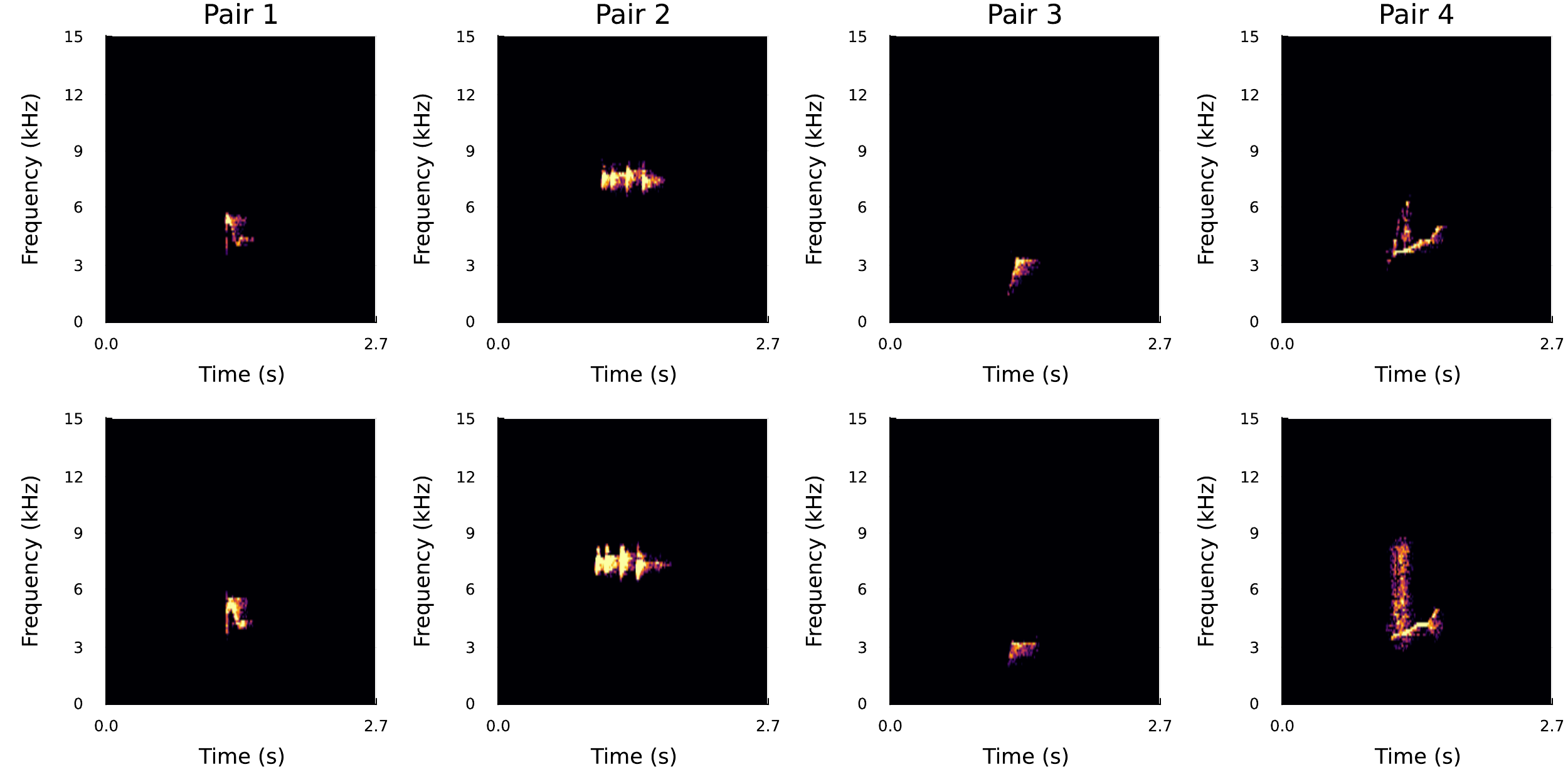}
  \caption{Examples of positive TFE pairs used to train the contrastive network. Pairs are obtained by associating the same sound event on two nearby microphones, allowing propagation effects to create natural ``augmentations".}
  \label{fig:tfe-pairs}
\end{figure*}

We train the embedding network using self-supervised contrastive learning. In traditional contrastive learning~\cite{chen2020simple}, positive pairs are created by applying different augmentations to the same training example, and the network is trained to place positive pairs close together while separating unrelated examples. We follow this overall framework, but introduce two modifications tailored to passive acoustic monitoring. Other than a random translation necessary to obtain TFEs of constant duration (Section~\ref{sec:tfe}), we do not perform synthetic augmentation. Instead, positive pairs are derived from recordings of the same acoustic event on the two microphones connected to each recorder unit, which are spaced approximately 10 meters apart (Figure~\ref{fig:sbg_sites}). The natural variability in sound propagation thus provides a source of ``augmentation" without requiring synthetic transformations. Consequently, paired TFEs can differ in both SNR and time-frequency structure (Figure~\ref{fig:tfe-pairs}). Since the sound may arrive at the two microphones at different times, the corresponding events on both microphones must first be associated. This association is performed using a combination of time information and a requirement for high cross-correlation between the two acoustic waveforms. These pairs may include any sound captured by the TFE segmentation---not exclusively bird vocalizations. The resulting representation space is referred to as the \emph{embedding space} hereafter. Because the approach is self-supervised, no class labels are used during this stage of training. The training is performed over 50~epochs using the Adam optimizer, but with a loss function that differs from the one proposed by Chen et al.~\cite{chen2020simple}.

Although the loss function proposed by Chen et al.~\cite{chen2020simple} is theoretically sound, it was developed for a computer vision task quite different from our needs and did not perform well on our problem. Based on the intuition that the embedding space can be modeled as a surface of a 1024-dimensional hypersphere, we do not seek maximal angular separation between dissimilar sounds, but rather orthogonality. Maximal angular separation pushes dissimilar sounds to diametrically opposite sides of the hypersphere, but that can only accommodate two classes. Instead, orthogonality pushes dissimilar sounds to different axes of the hypersphere, and can support up to 1024 distinct classes for a 1024-dimensional hypersphere. With this in mind, the loss function we use is:

\begin{align}
    &L(\mathbf{Z}) &=& \sum_{p=1}^N l_{2p,2p-1} + l_{2p-1,2p}, \label{eq:loss1} \\
    &l_{i,j} &=& -\left[ \mathbf{z}_i^T\mathbf{z}_j + \beta \sum_{k \ne i, k \ne j}^{2N} \min\left( 1 - \mathbf{z}_i^T\mathbf{z}_k, \; 1 \right)^2 \right]. \label{eq:loss2}
\end{align}

Here $\mathbf{Z} = \{\mathbf{z}_i \forall i\}$ is the set of embedding space representations for the training TFEs, organized such that the first pair is $(\mathbf{z}_1,\mathbf{z}_2)$, the second pair is $(\mathbf{z}_3,\mathbf{z}_4)$, and so on. The two terms in~(\ref{eq:loss1}) correspond to two possible orderings of samples in each pair. The first term in~(\ref{eq:loss2}) maximizes the similarity within each pair, while the second term induces orthogonality for non-paired representations. The $\min(\cdot, 1)$ function prevents maximal angular separation and $\beta = 3$ is a hyperparameter that controls the balance between the two terms, chosen empirically without extensive tuning. In practice, the loss is not evaluated over the entire dataset, but in mini-batches of $N=512$.

After training, we observe an average cosine similarity of 0.93 for positive pairs, while non-paired samples are approximately orthogonal. While this ensures that paired samples (similar sounds) have high similarity, it does not guarantee low similarity between every pair of dissimilar sounds. Since the loss function minimizes the \textit{average} similarity between dissimilar sounds, some clusters of dissimilar sounds still experience similar representations. This is addressed by allowing the embedding space representation to improve further in a final, supervised classification stage.

\subsubsection{Supervised classification}
\label{sec:classifier}

\begin{figure*}[t]
  \includegraphics[width=\linewidth]{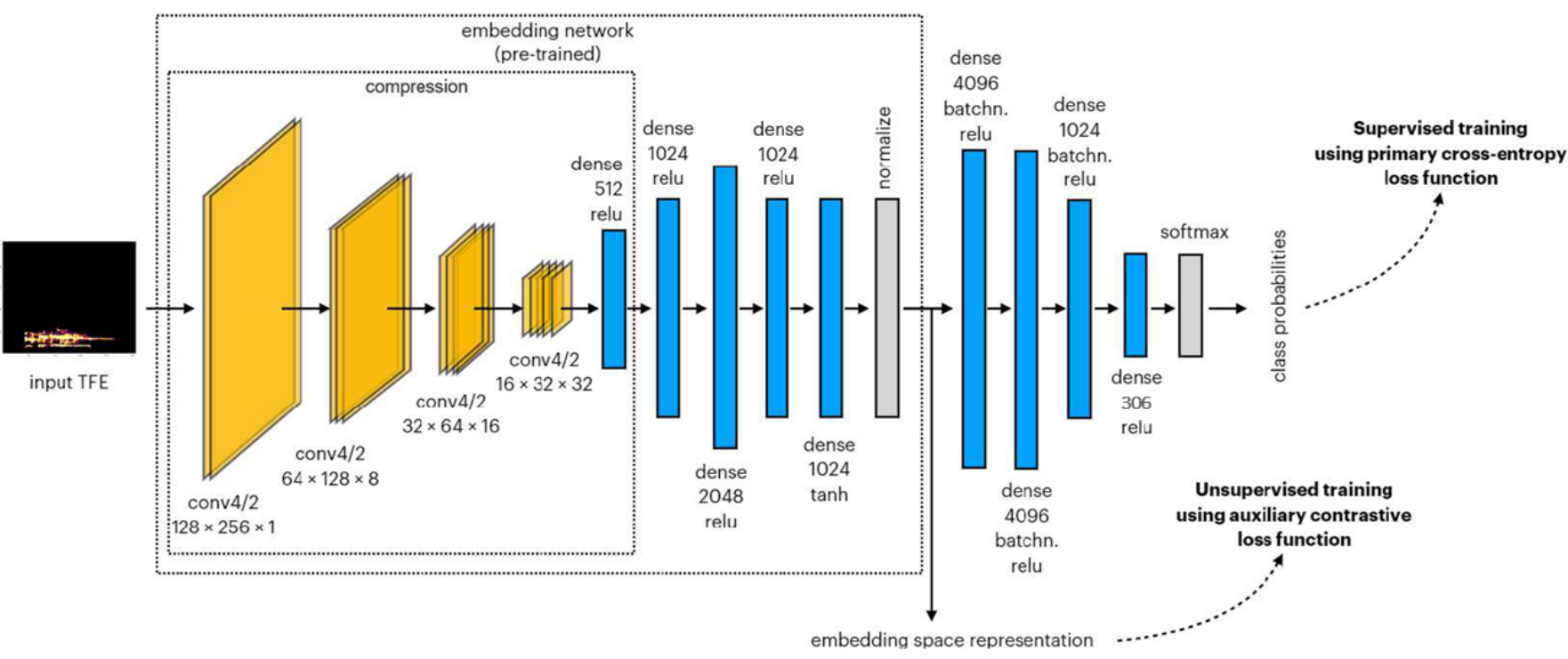}
  \caption{Complete LCC architecture. TFEs extracted from raw recordings are mapped to embedding vectors using the pre-trained contrastive network, followed by a four-layer classification head that outputs class confidence scores. The embedding network and classifier are trained in alternating rounds to ensure that the embeddings retain their contrastive objective while adapting to the classification task: five epochs of supervised end-to-end training of the full model (i.e., the embedding network is not frozen) are followed by one epoch of self-supervised contrastive training of the embedding network.}
  \label{fig:classifier}
\end{figure*}

The final step of the LCC method is a classification stage consisting of four dense layers with batch normalization between each layer. It is trained with a standard cross-entropy loss,takes the preceding TFE embedding as input and assigns a confidence score to each class (Figure~\ref{fig:classifier}). Classes are defined at call level rather than species level, by labeling groups of similar TFEs (e.g., Olive-winged Bulbul: call type 1). Predictions at species level, when needed, are obtained simply by ignoring the call-type distinctions.

To address the challenge that some clusters of dissimilar sounds occur close in embedding space (Section~\ref{sec:clr}), the classifier is trained end-to-end without freezing any preceding layers---allowing the embedding network to improve further. During this training, however, the embedding network can ``forget" that paired samples should have high similarity and unpaired samples low similarity. To reinforce this property, we simultaneously train the embedding network with the same contrastive loss as before (Section~\ref{sec:clr}). In practice, the training is implemented in sets of 5 epochs of the primary cross-entropy loss, followed by 1 epoch of the auxiliary contrastive loss. We train the network over 100 such sets and retain a mean similarity score of 0.866 for paired samples and near orthogonality (0.046) for unpaired samples.

\subsubsection{Extensions for long-duration deployments}
\label{sec:lcc-extensions}

The classifier is optimized to separate the training classes by minimizing the supervised loss. Therefore, it primarily learns decision boundaries tailored to the training distribution. When applied to continuous field recordings, however, it is exposed to a substantially broader range of acoustic events, many of which may produce embeddings that lie close to known classes and lead to false positives. To improve robustness in this setting, we introduce two extensions.

First, we train a separate \textit{bird-pass filter} to distinguish bird sounds from non-bird sounds. This binary classifier consists of three fully connected layers and operates on TFE embeddings as input. Since even a small false-positive rate can accumulate to many errors in long-duration recordings, we prioritize precision over recall. To enforce this, false positives (non-bird sounds classified as birds) are penalized more strongly than false negatives.

Second, we introduce an additional ``sink'' class in the primary classifier. This class captures non-bird sounds that are acoustically similar to bird vocalizations and therefore prone to being misclassified. It is trained using 1{,}893 TFEs from SBG site~\#1 that caused false positives when applying the LCC to continuous data. This encourages the network to learn discriminative features for such sounds while also providing an explicit rejection class for inputs that do not correspond to any of the target bird classes. To further reduce false positives in long-duration recordings, the sink class is assigned a higher loss weight. Specifically, as the sink class contains approximately 38 times more samples than each bird class, it is weighted by a factor of $3/38$---effectively weighting it three times more than each bird class.

\subsection{Fine-tuned BirdNET}
\label{sec:ftbn}

BirdNET is a widely used bioacoustic classifier first introduced in 2021~\cite{KAHL2021}, covering over 6{,}000 bird species as well as a small number of non-bird classes including human speech. We start from a pre-trained BirdNET~v2.4~\cite{birdnet_analyzer_git} model (BN-PT) and fine-tune it on data from Singapore (BN-FT) under two input schemes: local and global context. Local-context inputs have limited and variable time-frequency extent, similar to TFEs, while global-context inputs are fixed-duration broadband spectrograms, consistent with BirdNET's original training format. To allow adaptation of BN-FT to our data, we remove its original classifier head, freeze the backbone, and add a new trainable head with 306 outputs for local-context classification and 103 outputs for global-context classification. The architecture of the new head and other hyperparameters are obtained through BirdNET's autotuning feature. We run 20 hyperparameter trials on global context inputs with 5 runs per trial to reduce variance and select the configuration with the best performance on the validation set (Table~\ref{tab:hyperparameters}). The resulting classifier head has two dense layers with batch normalization between layers and ReLU activation functions. Hence, both BN-FT and the LCC have classifier heads with non-linear capacity.

\begin{table}[ht]
\caption{\label{tab:hyperparameters}Non-default BirdNET hyperparameters used to fine-tune BirdNET on both local- and global-context inputs. With the exception of the minimum frequency and the minimum confidence score, which are kept fixed for both models, the parameters are obtained in a hyperparameter search over 20 trials and 5 runs per trial using BirdNET's autotune option. The output dimension matches the number of target classes: 306 vocalization classes for local context and 103 species for global context.}
\begin{ruledtabular}
\begin{tabular}{lcc}
 & \multicolumn{2}{c}{\textbf{Input context}} \\
\cline{2-3}
\textbf{BirdNET parameter} & \textbf{Local} & \textbf{Global} \\
\hline
Output dim & 306 & 103\\
Hidden units & 256 & 512\\
Dropout & 0.33 & 0.5\\
Batch size & 64 & 64\\
Learning rate & 0.001 & 0.001\\
Upsampling ratio & 0.5 & 0.33\\
Label smoothing & Yes & No\\
Minimum frequency (Hz) & 500 & 500\\
Minimum confidence & 0.1 & 0.1\\
\end{tabular}
\end{ruledtabular}
\end{table}

\subsubsection{Fine-tuning on local context}
To compare the LCC with BirdNET, we fine-tune BirdNET on the same training data used for the LCC. However, BirdNET takes 3-second audio segments as input and is not directly compatible with either TFEs or TFE embeddings. To approximate the information content of TFEs, source recordings are first bandpass-filtered and time-cropped around each TFE\footnote{Although TFEs are defined in Section~\ref{sec:tfe} as matrices of size $128 \times 256$, only the bounding box around the non-zero energy regions are retained.}, and the resulting localized segments are embedded into 3-second clips of ambient soundscape recordings, with random time shifts applied to encourage time invariance. The ambient clips were extracted from recordings at SBG site~\#1 between 5 and 6~am, a period of low vocal activity, retaining only those confirmed to be free of vocalizations. Since ambient noise levels at this time are considerably lower than during daytime hours---when most TFEs are extracted---local-context inputs tend to have higher SNRs than their global-context counterparts, described next.

\subsubsection{Fine-tuning on global context}
Global-context inputs are obtained by extracting 3-second segments centered on the midpoint of each TFE, with frequencies limited to 0.5--15~kHz to match the range used during TFE extraction. Broader context may aid classification by capturing additional temporal and spectral cues relevant to the target species, but at the cost of also including vocalizations from other species and background noise. Unlike the local-context model, this variant is trained at species level, as the wider input window may capture multiple calls from the same species. As labels are tied to individual TFEs and are sparse and non-exhaustive, each sample is assigned a single ground-truth label, even if other species are audible.

\subsection{Dual-Context Classifier}
\label{sec:dcc}

The Dual-Context Classifier (DCC) leverages both local and global context by combining predictions from the LCC and BN-FT. The confidence scores of the two models are first normalized to account for differences in baseline confidence levels: for each model $m \in \{\text{LCC}, \text{BN-FT}\}$, the raw confidence score $\hat{y}_m$ is normalized as

\begin{equation}
\bar{\hat{y}}_m = \frac{\hat{y}_m}{\mu_m}
\end{equation}

where $\bar{\hat{y}}_m$ denotes the vector of normalized class confidence scores and $\mu_m$ is the mean confidence score among true positive predictions on the validation set.

We evaluate two fusion strategies: \emph{max voting} and \emph{weighted average voting}. In max voting, the final prediction is taken from the model with the highest normalized confidence score. In weighted average voting, normalized confidence scores are combined class-wise:

\begin{equation}
\hat{y}_{\text{DCC}}
=
\arg\max
\left(
w\bar{\hat{y}}_{\text{LCC}}
+
(1-w)\bar{\hat{y}}_{\text{BN-FT}}
\right)
\end{equation}

where the $\arg\max$ operator returns the class with the highest fused confidence score and $w=0.55$ controls the contribution of each model and is selected by maximizing the macro-averaged F1 score on the validation set.

\subsection{Datasets}
\label{sec:datasets}

We use three main datasets for training and evaluation (Table~\ref{tab:datasets}): (i) a self-supervised pre-training (SSPT) dataset used to train the embedding network, (ii) a supervised fine-tuning (SFT) dataset used to train the LCC, fine-tune BirdNET and evaluate in-distribution performance, and (iii) an out-of-distribution (OOD) dataset used to assess both LCC generalization and the effect of contrastive pre-training on downstream performance.

\subsubsection{Self-supervised pre-training dataset}

The SSPT dataset is constructed using the pairing strategy described in Section~\ref{sec:clr}, where simultaneous recordings from two microphones are used to form positive TFE pairs corresponding to the same acoustic event. We apply this strategy to a subset of days from the 7-month recording period across the two SBG deployments, yielding 19{,}311 reliable TFE pairs. We use 18{,}811 pairs for training and 500 pairs for validation (Table~\ref{tab:splits}).

\subsubsection{Supervised fine-tuning dataset}

The SFT dataset is curated from two sources: focal recordings from the Xeno-canto database~\cite{xenocanto} recorded in Singapore (63.5\%), and soundscape recordings from the two SBG sites (36.5\%). TFEs are first extracted from raw audio and converted to embeddings using the pre-trained embedding network. The resulting embeddings are clustered, and groups of similar TFEs are assigned labels by an avian expert based on representative samples. This procedure demonstrates how learned embeddings can be used as a tool for vocalization discovery, enabling the identification of acoustically similar sound groups directly from raw soundscape recordings, and thereby facilitating efficient annotation.

The curated SFT dataset contains 5{,}174 labeled TFEs spanning 306 vocalization classes across 103 bird species. The dataset is highly imbalanced, with class sizes ranging from a few samples to several hundred. We randomly assign samples to the training, validation, and test splits, ensuring that each class is represented by at least one sample in each split. To improve class balance during training, we cap the number of training samples per class at 50 and augment underrepresented classes using small random frequency shifts, minor time stretching, and additive Gaussian and white noise. Low-sample classes in the validation set are also augmented but capped at 20 samples per class. The test set, however, remains unaugmented to ensure final evaluation entirely on natural vocalizations (Table~\ref{tab:splits}). To avoid data leakage from the contrastive pre-training stage, we ensure that TFEs from the SSPT dataset are not present in the validation and test splits.

\subsubsection{Out-of-distribution dataset}
\label{sec:ood-dataset}
The OOD dataset is used to evaluate generalization of the Singapore-trained LCC model to new geographic regions and recording conditions. It is derived from the Xeno-canto subset of BirdSet~\cite{BirdSet}, the only BirdSet subset sharing species with the SFT dataset. We retain only recordings collected outside Singapore, resulting in 11{,}092 recordings covering 95 of the 103 species present in the SFT dataset.

Species labels are provided at the recording level, meaning that precise temporal annotations of vocalizations are not available. Recording durations vary, with a median length of 24 seconds. The number of recordings per species ranges from 5 to 1{,}064, with a mean of 117.

\subsubsection{Bird-pass filter dataset}

In addition to the three main datasets, we construct a separate dataset to train the bird-pass filter used in the long-duration deployment extension (Section~\ref{sec:lcc-extensions}). This dataset consists of two classes only: a bird class (positives) and a non-bird class (negatives). The positive class comprises all 5{,}174 non-augmented SFT samples. The negative class consists of 3{,}152 non-bird TFEs drawn primarily from the UrbanSound8K dataset~\cite{UrbanSound8k} and supplemented with underwater recordings of marine mammal vocalizations collected in Singapore waters~\cite{Hexeberg2023}. To balance class sizes, non-bird samples are augmented using random frequency shifts to exactly match the sample size of the bird class.

\begin{table*}[t]
\caption{\label{tab:datasets}Overview of three datasets used in this work: Self-supervised pre-training (SSPT), Supervised fine-tuning (SFT), and Out-of-distribution (OOD). TFE pairs for contrastive pre-training are derived from simultaneous recordings of the same vocalization on two microphones. The SFT set comprises a mixture of soundscape (36.5\%) and focal (63.5\%) recordings. Non-SG denotes recordings collected outside Singapore.}
\begin{ruledtabular}
\begin{tabular}{lllrllcc}
\textbf{Dataset} & \textbf{Source} & \textbf{Rec.\ type} & \textbf{Size} & \textbf{Sample type} & \textbf{Location} & \textbf{\#Classes} & \textbf{\#Species} \\
\hline
SSPT  & Self-collected      & Soundscape & 19{,}311 & TFE pairs  & Singapore & ---  & ---  \\
SFT   & Self-collected / XC & Mixed      &  5{,}174 & TFEs       & Singapore & 306  & 103  \\
OOD   & XC            & Focal      & 11{,}092 & Recordings & Non-SG  &  95  &  95  \\
\end{tabular}
\end{ruledtabular}
\end{table*}

\begin{table}[t]
\caption{\label{tab:splits}Data splits for self-supervised pre-training (SSPT) and supervised fine-tuning (SFT) datasets, with and without augmentation. The SFT test set remains unaugmented and consists exclusively of natural vocalizations, ensuring evaluation on real-world data.}
\begin{ruledtabular}
\begin{tabular}{llrrrr}
\textbf{Dataset} & \textbf{Aug.} & \textbf{Train} & \textbf{Val.} & \textbf{Test} & \textbf{Total} \\
\hline
SSPT & No  & 18{,}811 &    500 & ---     & 19{,}311 \\
\hline
SFT   & No  &  3{,}259 & 971 & 943 &  5{,}174 \\
SFT   & Yes & 15{,}300 & 6{,}257 & --- & 21{,}557 \\
\end{tabular}
\end{ruledtabular}
\end{table}

 \subsection{Evaluation metrics}
 \label{sec:metrics}

\subsubsection{Alignment of single- and multi-label outputs}
A key difference between BirdNET and the LCC lies in their output structure. BirdNET is multi-class multi-label, applying a binary classifier to each class, whereas the LCC is multi-class single-label, applying a single softmax across all classes. As a result, BirdNET can produce multiple high-confidence predictions for the same input---even if only a single species is present.

To ensure comparability between the multi-label and single-label approaches, we mark a prediction as correct if: (i) it matches the true class, (ii) has the highest confidence score among predictions, and (iii) its confidence exceeds 0.1. BirdNET occasionally assigns the maximum confidence score of 1.0 to multiple classes, resulting in ties. In such cases, one of the tied classes is selected at random. This tie-breaking rule prevents artificially inflated performance when multiple classes receive identical maximum scores.

\subsubsection{F1 metrics}
For the in-distribution SFT dataset, we report performance using micro-averaged (F1-micro) and macro-averaged (F1-macro) F1 scores. F1 is defined as the harmonic mean of precision and recall (Eq.~\ref{eq:f1}). F1-micro is computed by aggregating true positives (TP), false positives (FP), and false negatives (FN) across all $N$ samples (Eq.~\ref{eq:f1_micro}), whereas F1-macro is obtained by averaging class-wise F1 scores over the $C$ classes (Eq.~\ref{eq:f1_macro}).

\begin{equation}
\label{eq:f1}
F1 = 2 \cdot \frac{\text{Precision} \cdot \text{Recall}}{\text{Precision} + \text{Recall}}
\end{equation}

\begin{equation}
\label{eq:f1_micro}
\text{F1-micro} = 2 \cdot \frac{P_{\text{micro}} \cdot R_{\text{micro}}}{P_{\text{micro}} + R_{\text{micro}}}
\end{equation}
where
\begin{align*}
P_{\text{micro}} &= \frac{\sum_{i=1}^{N} \text{TP}_i}{\sum_{i=1}^{N} (\text{TP}_i + \text{FP}_i)} \\
R_{\text{micro}} &= \frac{\sum_{i=1}^{N} \text{TP}_i}{\sum_{i=1}^{N} (\text{TP}_i + \text{FN}_i)}
\end{align*}

\begin{equation}
\label{eq:f1_macro}
\text{F1-macro} = \frac{1}{C} \sum_{c=1}^{C} F1_c
\end{equation}

Both averaging schemes have limitations on the imbalanced SFT test set. F1-macro is sensitive to classes with very few samples whereas F1-micro is dominated by the performance of a small number of large classes. Therefore, performance should be interpreted jointly using both metrics.

To quantify the uncertainty associated with evaluating performance on a finite test set, we estimate 95\% confidence intervals for both F1 metrics using bootstrap resampling. Specifically, we generate 10{,}000 resampled test sets by sampling with replacement while preserving the original sample size, recompute the evaluation metrics for each resampled test set, and report the 2.5th and 97.5th percentiles.

\subsubsection{Hit rate}

The OOD dataset is annotated at the recording level, i.e., the exact time-frequency location(s) of the target species within each recording is unknown. In addition, most recordings contain vocalizations from other non-annotated background species. To evaluate performance under these conditions, we define a \textit{hit rate} metric. For each (recording, species) pair, a pair is counted as a hit if at least one predicted TFE matches the ground-truth species label. The hit rate is then computed as the fraction of recordings for which at least one hit is obtained.

\section{Results}

This section presents the experimental results. We begin by evaluating all classifiers on the in-distribution supervised fine-tuning (SFT) test set from Singapore (Section~\ref{sec:sft_results}). We then assess the Local-Context Classifier (LCC) under two additional conditions: out-of-distribution recordings collected outside Singapore (Section~\ref{sec:ood_results}) and continuous soundscape recordings in a deployment-like setting (Section~\ref{sec:continuous_results}). Next, we summarize the performance of the time-frequency event (TFE) extraction stage based on a separate study (Section~\ref{sec:tfe_results}). Finally, we analyze the effect of contrastive pre-training on downstream classification performance (Section~\ref{sec:contrastive_results}).

\subsection{In-distribution recordings}
\label{sec:sft_results}

For evaluation on the in-distribution SFT test set, we group methods by their input context: local context, global context, and combined local and global context. Within the local-context category, we compare the proposed LCC against two BirdNET variants: the original pre-trained model (BN-PT) and a version fine-tuned on local data (BN-FT). We additionally evaluate an ensemble of LCC and BN-FT, constructed analogously to the DCC (Section~\ref{sec:dcc}) except that the models both operate on local-context inputs. For the global-context category, we only evaluate BirdNET models as the LCC method is inherently local-context. Finally, the DCC combines the LCC with the global-context BN-FT model and is the only method that leverages both context types.

The BN-PT model is evaluated without additional training, while all BN-FT models are trained on the same training samples as the LCC, differing only in context and input representation (Section~\ref{sec:ftbn}). Since BirdNET was originally trained on more than 6{,}000 species, while our test set covers only 103, we restrict BN-PT to these 103 species to ensure a fair comparison with LCC. 

We find that BirdNET exhibits some performance variability across training runs. To reduce the influence of this variability, each BN-FT configuration is trained 10 times for 50 epochs using the tuned hyperparameters in Table~\ref{tab:hyperparameters}. The model achieving the highest validation-set performance is then selected for test-set evaluation.

The resulting test-set performance is summarized in Table~\ref{tab:test-set-comparison}. The LCC outperforms the pre-trained BirdNET models on both local and global contexts. BirdNET improves substantially after fine-tuning on in-distribution data in both contexts, highlighting the challenge of generalization even for models pre-trained on large, diverse datasets.

While local context allows time-frequency localization of calls, the global context results suggest that larger context improves classification performance. Combining local and global context in the DCC improves performance on the test set further, suggesting that local and global representations provide complementary information. Using weighted average voting ($w=0.55$), the DCC achieves F1-micro and F1-macro scores of 94.6\% and 93.8\%, respectively. These correspond to gains of 6.9 and 1.7 percentage points over the best-performing BirdNET model. This ensemble strategy also yields higher performance than max voting, which achieves F1 micro and macro scores of 93.2\% and 91.8\%, respectively.

\begin{table*}[ht]
\caption{\label{tab:test-set-comparison}Performance on the SFT test set. Values are reported as F1 scores with 95\% bootstrap confidence intervals. LCC operates on local-context TFE inputs, whereas BirdNET is evaluated on both local- and global-context inputs. Models operating on local context are evaluated at the TFE level (except BN-PT), and all other models at the species level.}
\begin{ruledtabular}
\begin{tabular}{llcc}
\textbf{Input context} & \textbf{Model} & \textbf{F1-micro} & \textbf{F1-macro} \\
\hline
\multirow{4}{*}{Local}
& LCC & 0.793 (0.767--0.819) & 0.685 (0.663--0.730) \\
& BN-PT & 0.265 (0.238--0.294) & 0.296 (0.253--0.326) \\
& BN-FT & 0.780 (0.752--0.805) & 0.717 (0.699--0.766) \\
& LCC + BN-FT & 0.883 (0.861--0.902) & 0.820 (0.797--0.858) \\
\hline
\multirow{2}{*}{Global}
& BN-PT & 0.619 (0.587--0.650) & 0.622 (0.603--0.680) \\
& BN-FT & 0.877 (0.852--0.894) & 0.921 (0.900--0.942) \\
\hline
Local + Global & DCC & 0.946 (0.930--0.960) & 0.938 (0.919--0.963) \\
\end{tabular}
\end{ruledtabular}
\end{table*}

\subsection{Out-of-distribution recordings}
\label{sec:ood_results}

The OOD dataset consists of 11{,}092 weakly labeled Xeno-canto recordings from 95 species recorded outside Singapore (Section~\ref{sec:ood-dataset}). BirdNET is not evaluated on this dataset because its pre-training data includes recordings from Xeno-canto.

\begin{figure}[t]
\centerline{\includegraphics[width=\linewidth,keepaspectratio]{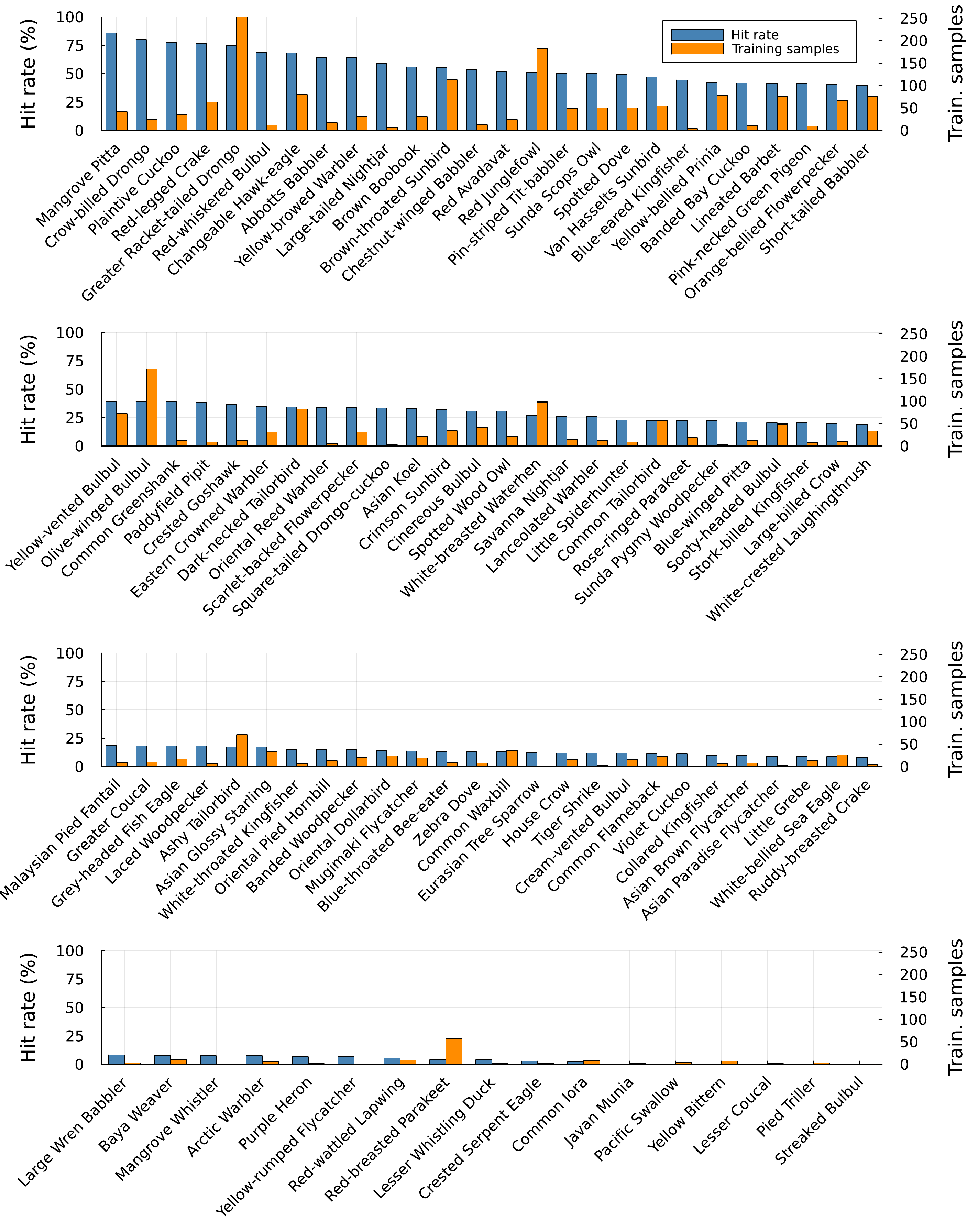}}
    \caption{LCC performance by species on the OOD dataset showing hit rates (blue) and the number of non-augmented training samples for the corresponding species (orange).}
    \label{fig:hitrate-ood}
\end{figure}

At the recording level, the LCC detects the ground-truth species in 3{,}084 of the 11{,}092 recordings, corresponding to a micro hit rate of 27.8\%. Note that hit rate measures the proportion of recordings containing at least one correct detection but does not reflect detection frequency. Among recordings with at least one correct detection, the mean detection rate is 5.4 detections per minute across a combined 48.2 hours of audio. Species-level hit rates range from 0\% to 85.7\%, with a macro-average of 27.8\%. As expected, species-level performance generally increases with the number of training samples (Figure~\ref{fig:hitrate-ood}). For example, the six species with a 0\% hit rate are represented by only 1--7 TFEs prior to augmentation.

However, there are also species that achieve relatively strong performance despite limited training data, indicating that performance is influenced by factors beyond training set size. The Large-tailed Nightjar, for instance, is trained on only seven non-augmented TFEs from a single 15-second recording, yet is detected in 66 of 112 recordings, yielding a hit rate of 58.9\%. The total number of detections in these 66 recordings is 1{,}487, equivalent to 29.1 detections per minute across 51 minutes of audio. Of the 46 recordings without any correct detections, 17 contain a call type absent from the SFT training data, while the remaining 29 are misclassifications.

\subsection{Continuous soundscape recordings}
\label{sec:continuous_results}

 \begin{figure*}[!t]
  \includegraphics[width=\linewidth]{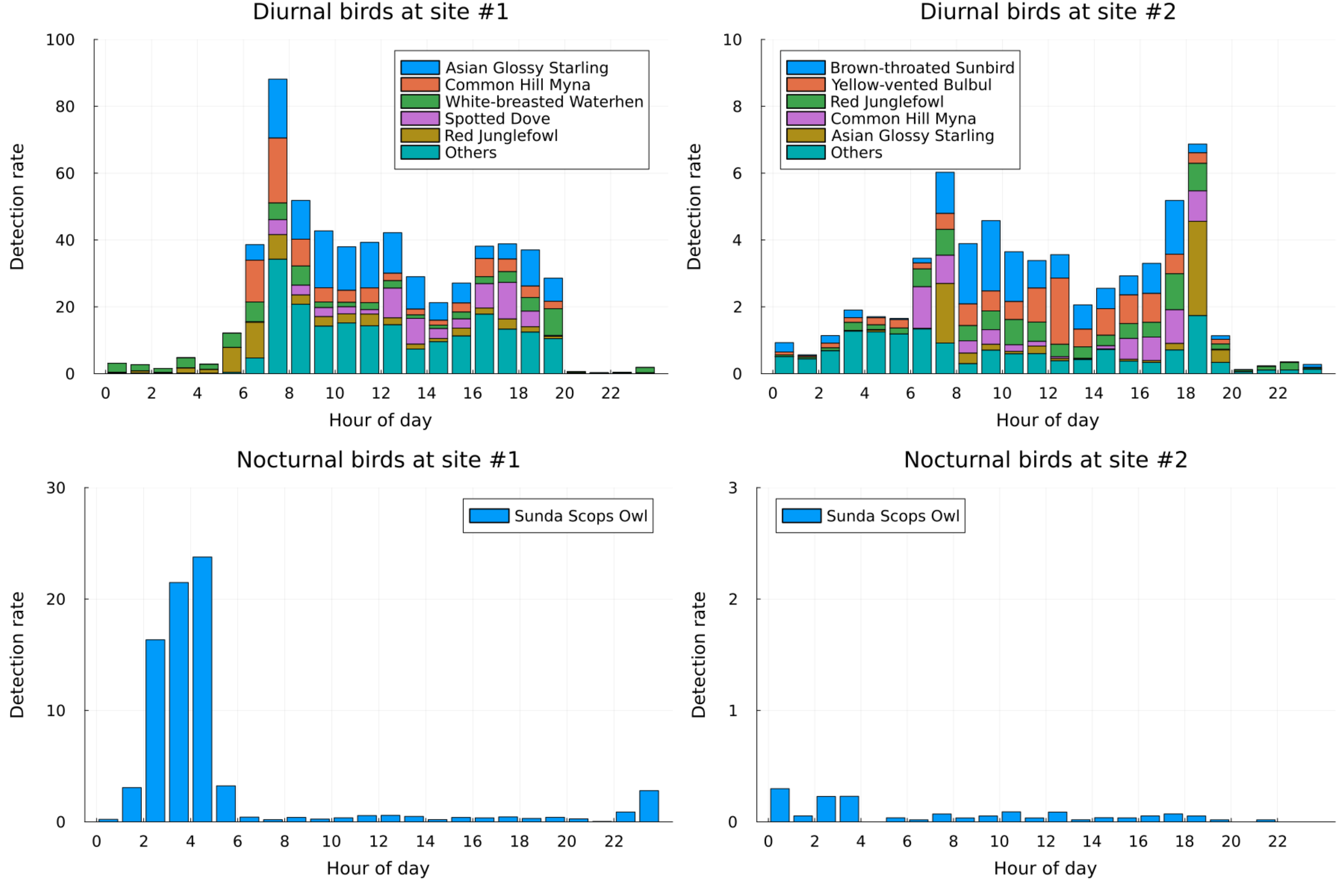}
  \caption{Hourly detection rates for the 16 shortlisted species, averaged over a 10-day period at both SBG sites and grouped into diurnal and nocturnal species. The five species with the highest detection counts at each site are listed by name. Note that the y-axis ranges differ between the sites by an order of magnitude.}
  \label{fig:diurnal_and_nocturnal_species_by_hour}
\end{figure*}

The SFT test set evaluates the detector on isolated vocalizations from species represented during training. In a practical deployment, however, the detector is applied to continuous recordings where many extracted TFEs correspond to non-bird sounds or vocalizations outside the model's target classes. To assess feasibility in such conditions, we conduct a preliminary evaluation on continuous soundscape recordings from the SBG. The objectives are twofold: (i) to estimate performance for a subset of promising classes in a realistic long-term monitoring scenario, and (ii) to demonstrate how detector outputs can be aggregated to reveal ecologically meaningful temporal activity patterns. We limit the evaluation to precision, as estimating recall at the vocalization level from continuous recordings would require substantial manual annotation effort.

To keep the analysis tractable, we restrict the evaluation to a subset of classes. We first exclude species known to be absent from the SBG and migratory species not expected during the evaluation period, analogous to the geo-temporal filtering available in frameworks such as BirdNET~\cite{birdnet_analyzer_git}. From the remaining classes, we select vocalization classes that exhibit characteristic vocalizations suitable for manual verification. Following manual verification, only classes with sufficient precision are retained for the subsequent long-term analysis. This selection introduces a positive bias toward higher-performing classes; consequently, the reported precision and temporal activity patterns should be interpreted as a focused assessment of deployment-ready classes rather than an estimate of overall performance across all 306 classes.

The LCC is applied to one full day of recordings from SBG site~\#1. Following bird-pass filtering at a threshold of 0.5 (Section~\ref{sec:lcc-extensions}), up to 50 high-confidence predictions (confidence $\geq 0.7$) are randomly selected from each class for manual verification, yielding 958 samples in total. Of these, 732 are true positives and 226 are false positives, corresponding to a precision of 76.4\% when averaged over all samples and 78.1\% when averaged over classes (Table~\ref{tab:precisionContinuousData}). Several classes achieve high precision despite limited training data, suggesting that distinctive vocalization types can be learned from relatively few annotated examples. For example, one Brown-throated Sunbird class is trained on only four non-augmented samples yet correctly classifies 41 of 50 verified predictions (82\%).

To illustrate long-term monitoring applications, the detector is subsequently applied to 10 days of recordings from each SBG site between July 2020 and January 2021. With six microphones recording in parallel (and occasional downtime), this translates to approximately 1350 microphone-hours of recordings per site. Figure~\ref{fig:diurnal_and_nocturnal_species_by_hour} shows hourly detection rates (detections per microphone-hour) averaged over the 10-day period for the shortlisted species. Detections from different months are naturally aligned with dawn and dusk because sunrise and sunset in Singapore remain close to 7~am and 7~pm throughout the year.

The resulting temporal patterns are largely consistent with expected bird behavior. Diurnal species exhibit pronounced peaks around dawn and dusk, with relatively few detections during nighttime. In contrast, the only nocturnal species among the shortlisted classes, the Sunda Scops Owl (\textit{Otus lempiji}), is detected primarily between 2 and 5 am. As nocturnal activity declines, the Red Junglefowl (\textit{Gallus gallus}) dominates the soundscape at site~\#1 between 5 and 6~am, consistent with its well-known pre-dawn calling behavior.

Detections outside the expected activity periods are likely false positives, consistent with the precision estimates in Table~\ref{tab:precisionContinuousData}. Site~\#2, in particular, exhibits a relatively high detection rate of White-breasted Waterhen (\textit{Amaurornis phoenicurus}) between midnight and dawn. Inspection of a sample of these nighttime detections indicates that most are false positives originating from low-frequency traffic sounds that generate TFEs resembling those of White-breasted Waterhen vocalizations. This highlights the importance of robust generalization across soundscapes, as even nearby survey locations can differ substantially in background noise characteristics.

\begin{table*}[ht]
\caption{\label{tab:precisionContinuousData}Performance of shortlisted classes from a full day of continuous soundscape recordings at site~\#1. Bird names follow the Checklist of Bird Species for Singapore~\cite{nparksBirdsSingapore}.}
\footnotesize
\setlength{\tabcolsep}{2pt}       
\begin{ruledtabular}
\begin{tabular}{@{}ccllccc@{}}
\textbf{Class} & \shortstack{\textbf{Training}\\\textbf{samples}} & \textbf{Common name} & \textbf{Scientific name} & \textbf{Detections} & \textbf{TPs} & \textbf{Precision} \\
\hline
1  & 33 & Asian Glossy Starling        & \textit{Aplonis panayensis}     & 50 & 44 & 0.88 \\
2  & 6  & Black-naped Oriole           & \textit{Oriolus chinensis}      & 14 & 10 & 0.71 \\
3  & 4  & Blue-eared Kingfisher        & \textit{Alcedo meninting}       & 50 & 30 & 0.60 \\
4  & 34 & Brown-throated Sunbird       & \textit{Anthreptes malacensis}  & 36 & 34 & 0.94 \\
5  & 4  & Brown-throated Sunbird       & \textit{Anthreptes malacensis}  & 50 & 41 & 0.82 \\
6  & 3  & Changeable Hawk-eagle        & \textit{Nisaetus cirrhatus}     & 2  & 2  & 1.00 \\
7  & 50 & Common Hill Myna             & \textit{Gracula religiosa}      & 50 & 42 & 0.84 \\
8  & 6  & Common Hill Myna             & \textit{Gracula religiosa}      & 24 & 22 & 0.92 \\
9  & 3  & Common Hill Myna             & \textit{Gracula religiosa}      & 1  & 1  & 1.00 \\
10 & 6  & Crimson Sunbird              & \textit{Aethopyga siparaja}     & 10 & 7  & 0.70 \\
11 & 1  & Greater Racket-tailed Drongo & \textit{Dicrurus paradiseus}    & 14 & 7  & 0.50 \\
12 & 11 & Olive-backed Sunbird         & \textit{Cinnyris jugularis}     & 34 & 33 & 0.97 \\
13 & 15 & Olive-backed Sunbird         & \textit{Cinnyris jugularis}     & 20 & 15 & 0.75 \\
14 & 50 & Olive-winged Bulbul          & \textit{Pycnonotus plumosus}    & 50 & 40 & 0.80 \\
15 & 22 & Olive-winged Bulbul          & \textit{Pycnonotus plumosus}    & 32 & 25 & 0.78 \\
16 & 30 & Olive-winged Bulbul          & \textit{Pycnonotus plumosus}    & 50 & 26 & 0.52 \\
17 & 33 & Olive-winged Bulbul          & \textit{Pycnonotus plumosus}    & 50 & 42 & 0.84 \\
18 & 6  & Red-whiskered Bulbul         & \textit{Pycnonotus jocosus}     & 50 & 25 & 0.50 \\
19 & 33 & Red Junglefowl               & \textit{Gallus gallus}          & 50 & 44 & 0.88 \\
20 & 26 & Red Junglefowl               & \textit{Gallus gallus}          & 50 & 33 & 0.66 \\
21 & 50 & Red Junglefowl               & \textit{Gallus gallus}          & 6  & 5  & 0.83 \\
22 & 50 & Spotted Dove                 & \textit{Spilopelia chinensis}   & 50 & 48 & 0.96 \\
23 & 50 & Sunda Scops Owl              & \textit{Otus lempiji}           & 50 & 37 & 0.74 \\
24 & 18 & White-breasted Waterhen      & \textit{Amaurornis phoenicurus} & 15 & 12 & 0.80 \\
25 & 11 & White-breasted Waterhen      & \textit{Amaurornis phoenicurus} & 50 & 28 & 0.56 \\
26 & 26 & Yellow-vented Bulbul         & \textit{Pycnonotus goiavier}    & 50 & 36 & 0.72 \\
27 & 11 & Yellow-vented Bulbul         & \textit{Pycnonotus goiavier}    & 50 & 43 & 0.86 \\
\end{tabular}
\end{ruledtabular}
\end{table*}

\subsection{TFE extraction performance}
\label{sec:tfe_results}

As the first stage of the pipeline, TFE extraction directly affects all downstream stages. Vocalizations missed at this stage cannot be recovered in later stages, while TFEs from non-target sources such as insects or anthropogenic noise must either be filtered by the bird-pass filter or assigned to the sink class to avoid becoming false positives. In the following, we summarize the TFE extraction performance reported in our previous work~\cite{hexeberg2026}.

The TFE method is evaluated on its ability to localize bird vocalizations (i.e., distinguishing bird sounds from non-bird sounds) in time and frequency. Performance is measured using Intersection over Minimum (IoMin), an overlap metric designed for acoustic events which is more tolerant of annotation ambiguities than conventional intersection-over-union (IoU). Such ambiguities arise because vocalizations often have diffuse temporal and spectral boundaries, and because annotators may disagree on whether temporally fragmented vocalizations should be labeled as a single event or multiple events~\cite{hexeberg2026}. IoMin is defined as

\begin{equation}
    \text{IoMin} =
    \frac{\text{intersection}}
         {\min(\text{area}_{TFE},\ \text{area}_{gt})},
    \label{eq:iomin}
\end{equation}

where $\text{area}_{TFE}$ and $\text{area}_{gt}$ denote the bounding box areas of the detected TFE and the ground-truth vocalization, respectively.

\begin{figure}[!t]
  \includegraphics[width=\linewidth]{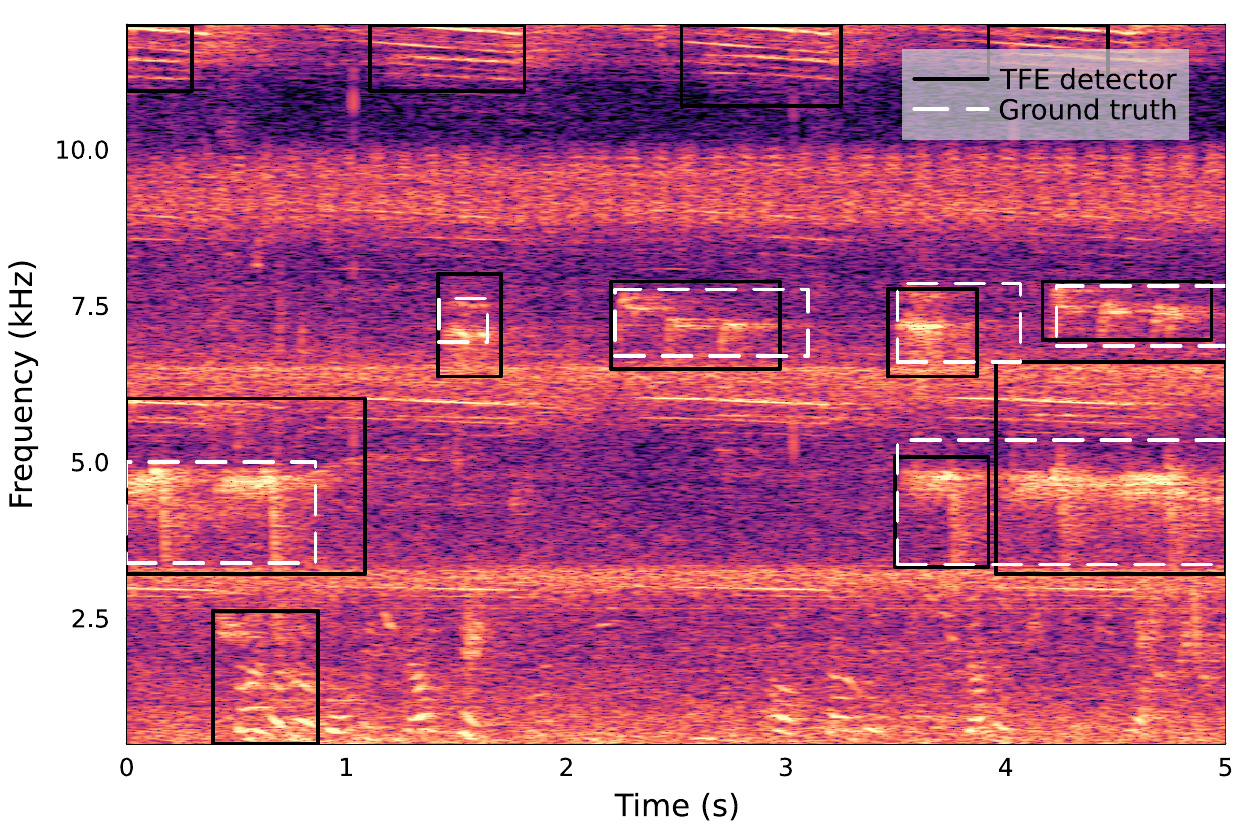}
  \caption{Example of extracted TFEs and human-labeled bird vocalizations (reproduced from \cite{hexeberg2026}). The TFE method detects both bird and non-bird acoustic events, including insect noise (top) and human speech (bottom).}
  \label{fig:tfe-example}
\end{figure}

Using a minimum overlap threshold of 50\% (IoMin@50) to declare a true positive, the method achieves an F1-score of 42.1\% on 18{,}095 labeled vocalizations from SBG and 48.6\% on 81{,}691 vocalizations across four sites in Hawaii. Two common limitations of the TFE method are illustrated in Figure~\ref{fig:tfe-example}: imperfect vocalization boundaries and false detections caused by non-bird sounds such as insect noise and human speech.

\subsection{Effect of contrastive pre-training}
\label{sec:contrastive_results}

To measure the effect of contrastive pre-training, we train an LCC model without it (LCC*) and compare it to the pre-trained LCC on both the in-distribution SFT and out-of-distribution (OOD) test sets. The two models are architecturally identical and differ only in the use of contrastive pre-training. Contrastive pre-training has a small negative effect on the in-distribution SFT test set, reducing the number of correctly classified TFEs by 13 (-1.7\% relative change). In contrast, performance on the OOD dataset improves substantially, with the number of recordings with hits increasing from 2{,}596 to 3{,}084, corresponding to an 18.8\% relative gain (Table~\ref{tab:ablation}).

\begin{table}[t]
\centering
\caption{Comparison of LCC and LCC without contrastive pre-training (LCC*) on the SFT and OOD test sets.}
\label{tab:ablation}
\begin{tabular}{lccccr}
\hline\hline
Dataset & Metric & Aggregation & LCC* & LCC & Rel.\ Gain \\
\hline
\multirow{2}{*}{SFT} & \multirow{2}{*}{Accuracy} & Micro & 82.4\% & 81.0\% & $-1.7\%$ \\
                     &                           & Macro & 74.4\% & 72.8\% & $-2.2\%$ \\
\hline
\multirow{2}{*}{OOD} & \multirow{2}{*}{Hit Rate} & Micro & 23.4\% & 27.8\% & $+18.8\%$ \\
                     &                           & Macro & 25.0\% & 27.8\% & $+11.2\%$ \\
\hline\hline
\end{tabular}
\end{table}

\section{Discussion}

\begin{figure*}[t]
\centerline{\includegraphics[width=\linewidth,keepaspectratio]{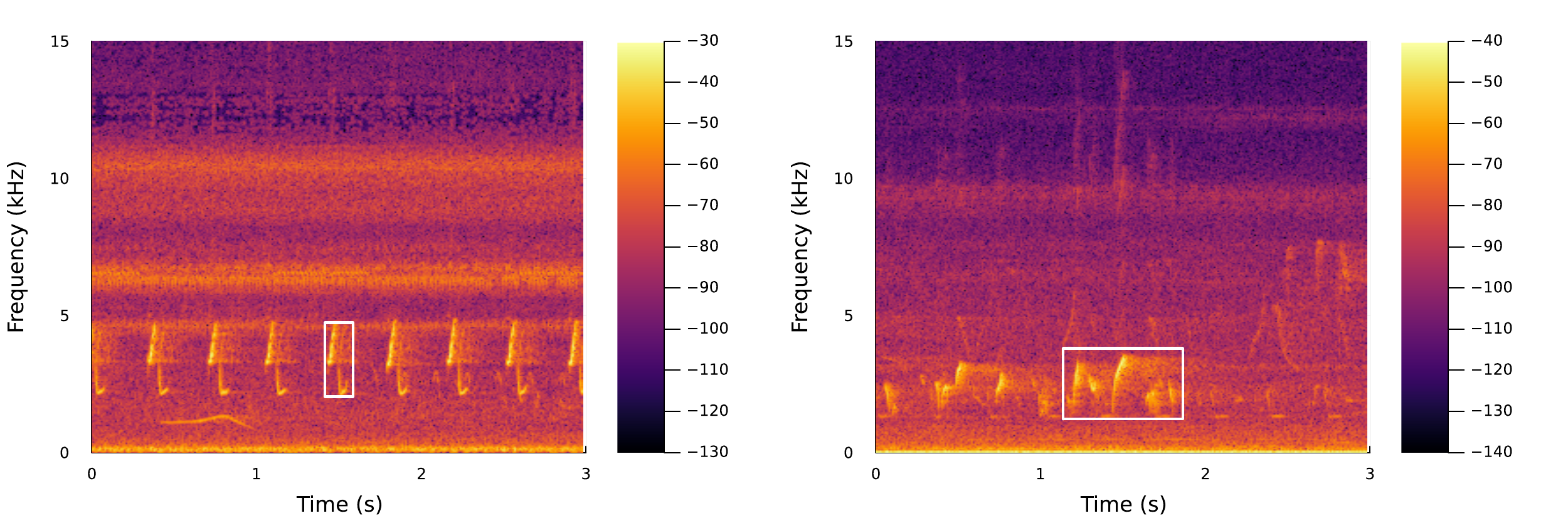}}
    \caption{Examples from the SBG recordings illustrating the complementary strengths of local and global context. \textbf{Left:} A repeating call sequence from a Common Tailorbird (\textit{Orthotomus sutorius}) together with a low-frequency call from an Asian Koel (\textit{Eudynamys scolopaceus}) around 0.5--1~s. Global-context models such as BirdNET may detect both species but cannot associate predictions with individual vocalizations, whereas the LCC preserves time-frequency localization. \textbf{Right:} A dominant call from an Olive-winged Bulbul (\textit{Pycnonotus plumosus}) together with several low-SNR background calls. Localized TFEs do not always capture complete vocalization sequences, illustrating how broader acoustic context can support classification.}
    \label{fig:local_vs_global}
\end{figure*}

Many bioacoustic studies aim to understand animal behavior in natural acoustic environments, including how vocal behavior adapts to conditions such as anthropogenic noise. Such studies therefore benefit from models that can both accurately classify species in noisy conditions and localize vocalizations in time and frequency. Singapore provides a useful testbed, combining high biodiversity with dense urbanization. The SBG, in particular, contains diverse acoustic sources---including birds, amphibians, reptiles, heavy rainfall, and human activity from nearby traffic and millions of annual visitors.

The evaluation results from Singapore (Table~\ref{tab:test-set-comparison}) suggest that local and global context provide complementary value: while local context enables source localization, global context provides additional information beneficial for species classification. Figure~\ref{fig:local_vs_global} illustrates these complementary roles. In the left example, two species vocalize within the same spectrogram window. Although both BN-FT and the LCC identify the correct species, only the LCC can associate predictions with individual vocalizations. In the right example, the LCC captures only part of the target call, whereas global context provides additional acoustic information that may improve classification.

The results also emphasize the importance of region-specific training data. Fine-tuning BirdNET on recordings from Singapore yields substantial performance gains, indicating that local adaptation remains important even for foundation models trained on large and diverse datasets. This presents a practical challenge for biodiversity monitoring, as collecting and annotating regional datasets can be costly. The challenge is particularly pronounced in settings where monitoring is most needed: biodiverse but remote habitats such as tropical rainforests, and for rare or endangered species, whose vocalizations are inherently difficult to capture in large quantities. This highlights the value of frameworks that are both sample-efficient and support efficient labeling.

The proposed LCC framework helps reduce dependence on annotated regional datasets in two ways. First, the embedding network is pre-trained entirely through self-supervised learning without manually labeled data. Second, because the embedding network is trained to group acoustically similar sounds, it can support efficient annotation of new recordings by clustering embeddings extracted from raw soundscape data, allowing groups of similar segments to be labeled in bulk rather than individually.

Consistent with this, the evaluations on continuous and out-of-distribution recordings (Table~\ref{tab:precisionContinuousData} and Figure~\ref{fig:hitrate-ood}) show that high performance can be achieved with very limited training data. Moreover, the contrastive pre-training not only enables efficient search in novel soundscapes but also improves generalization (Table~\ref{tab:ablation}).

While contrastive pre-training improves generalization on out-of-distribution data, a performance gap remains between in-distribution and out-of-distribution evaluation. Part of this gap reflects true domain shift, similar to the degradation observed when BirdNET is applied to Singapore recordings without local fine-tuning. However, two additional factors may contribute to the observed errors. First, although the LCC is trained on species present in the OOD dataset, the available training data may not capture the full diversity of vocalization types for each species. This results in cases where target calls in the OOD data are not represented in the training distribution; for the Large-tailed Nightjar in Section~\ref{sec:ood_results}, for example, 17 of 46 false negative recordings (37.0\%) arise from missing vocalization types rather than poor generalization. Second, the TFE method does not retain all bird vocalizations, as reported in Section~\ref{sec:tfe_results}. Consequently, some vocalizations are lost at this initial stage and cannot be classified downstream.

Together, these findings suggest several avenues for improving robustness and coverage in future work.

\section{Future directions}
\label{sec:further_work}

\subsection{Scale up training data}

Although the proposed method demonstrates that high performance can be achieved with limited training data, several classes remain weak, particularly in out-of-distribution settings. This likely reflects both limited sample size and incomplete coverage of species-specific vocalization types. A natural extension is therefore to scale training across three axes to improve robustness: dataset size, vocalization diversity, and geographic coverage.

\subsection{Improve TFE extraction}

The current TFE extraction stage identifies many bird vocalizations but cannot fully distinguish them from other high-energy, frequency-modulated signals such as insect noise or human speech. Because this is the first stage in the pipeline, its errors propagate downstream: non-bird TFEs can lead to false positives, while missed detections result in irrecoverable false negatives. Improving this stage is particularly important for long-duration passive acoustic monitoring where non-bird TFEs can accumulate in large numbers. One promising direction is to replace the current TFE extractor and bird-pass filter with a single, learnable bird vocalization detector, as proposed by Hexeberg et al.~\cite{hexeberg2026}.

\subsection{Combine time-frequency localization with global context}

While local context enables localization of calls in time and frequency, we find that broader context improves species classification. Exploring methods that leverage global context while still preserving time-frequency localization would therefore be an interesting direction for future work. The proposed dual-context classifier is a first step in this direction. While it achieves high classification accuracy, precise localization is only available through the LCC. Consequently, if the LCC fails to detect a target species while BN-FT succeeds, localization is still unavailable. A more integrated approach could instead treat TFE embeddings as tokens and model their temporal relationships using sequence-based architectures such as transformers, enabling joint modeling of local events and broader acoustic context.

\section{Conclusions}

In this work, we address a limitation of prevailing bioacoustic classifiers by introducing (i) a Local-Context Classifier that operates on localized time–frequency representations rather than broadband, fixed-duration spectrograms, and (ii) a Dual-Context Classifier that combines local and global context. By isolating individual vocalizations in both time and frequency, the framework simplifies the learning objective in complex soundscapes and supports studies where precise localization of calls is important, such as investigations of how animals adapt their vocalizations in response to changes in their acoustic environment.

We train and evaluate the methods on recordings from Singapore's dense and acoustically complex soundscape, which combines high biodiversity with multiple sources of anthropogenic noise. Our results highlight two main findings. First, while localized inputs enable precise source localization, broader time-frequency context improves species classification: the DCC achieves the highest F1 scores among the tested methods on a test set comprising 103 bird species. Second, we show that strong performance can be achieved with limited training data. While there is growing interest in large-scale foundation models, our results indicate that fine-tuning on regional data remains important for achieving high performance. However, the cost of collecting labeled data in new regions highlights the need for data-efficient methods. The proposed contrastive pre-training strategy is entirely self-supervised and improves generalization on an out-of-distribution test set. The resulting embedding network can further support efficient annotation by clustering acoustically similar sounds in novel soundscapes.

Although validated on bird data, the proposed framework is not restricted to avian bioacoustics but can be applied to other acoustic monitoring tasks involving clustering or classification of frequency-modulated sounds. We hope this work not only helps bridge an important gap in bioacoustic soundscape analysis but also motivates further exploration of methods that integrate local and global time–frequency context in bioacoustic applications.

\begin{acknowledgments}
This research was supported by The National Parks Board – Singapore (NParks).
\end{acknowledgments}

\bibliographystyle{jasanum2}
\bibliography{references}

@article{KAHL2021,
  title={BirdNET: A deep learning solution for avian diversity monitoring},
  author={Kahl, Stefan and Wood, Connor M and Eibl, Maximilian and Klinck, Holger},
  journal={Ecological Informatics},
  volume={61},
  pages={101236},
  year={2021},
  publisher={Elsevier}
}

@inproceedings{chen2020simple,
  title     = {A Simple Framework for Contrastive Learning of Visual Representations},
  author    = {Chen, Ting and Kornblith, Simon and Norouzi, Mohammad and Hinton, Geoffrey},
  booktitle = {Proceedings of the 37th International Conference on Machine Learning (ICML)},
  series    = {Proceedings of Machine Learning Research},
  volume    = {119},
  pages     = {1597--1607},
  year      = {2020},
  publisher = {PMLR}
}

@INPROCEEDINGS{Hexeberg2023,
  author={Hexeberg, Simen and Vishnu, Hari and Beng, Koay Teong and Ho, Abel and Yusong, Wu and Chitre, Mandar and Tun, Karenne and Lim, Karen},
  booktitle={OCEANS 2023 - Limerick}, 
  title={Acoustic detector for multiple vocalizing marine mammal individuals}, 
  year={2023},
  volume={},
  number={},
  pages={1-8},
  doi={10.1109/OCEANSLimerick52467.2023.10244432}}

@ARTICLE{Stowell2022,
  title    = "Computational bioacoustics with deep learning: a review and roadmap",
  author   = "Stowell, Dan",
  journal  = "PeerJ",
  volume   =  10,
  pages    = "e13152",
  year     =  2022
}

@inbook{Sutherland1998,
author = {Sutherland, Louis and Daigle, Gilles},
year = {1998},
publisher = {J. Wiley},
pages = {341-365},
title = {Atmospheric Sound Propagation}
}

@book{bibby2000,
    author = {Bibby, Colin J. and Burgess, Neil D. and Hill, David A.},
    title = {Bird census techniques},
    publisher = {Academic Press},
    year = {2000},
    address = {London, UK},
    edition = {Revised}
}

@inproceedings{UrbanSound8k,
    Address = {Orlando, FL, USA},
    Author = {Salamon, J. and Jacoby, C. and Bello, J. P.},
    Booktitle = {22nd {ACM} International Conference on Multimedia (ACM-MM'14)},
    Month = {Nov.},
    Pages = {1041--1044},
    Title = {A Dataset and Taxonomy for Urban Sound Research},
    Year = {2014}}

@article{darras2018,
  title={Estimating bird detection distances in sound recordings for standardizing detection ranges and distance sampling},
  author={Darras, Kevin and Furnas, Brett and Fitriawan, Irfan. and Mulyani, Yeni and Tscharntke, Teja},
  journal={Methods in Ecology and Evolution},
  volume={9},
  pages={1928--1938},
  year={2018},
  url={https://doi.org/10.1111/2041-210X.13031}
}

@article{Darras2019,
author = {Darras, Kevin and Batáry, Péter and Furnas, Brett J. and Grass, Ingo and Mulyani, Yeni A. and Tscharntke, Teja},
title = {Autonomous sound recording outperforms human observation for sampling birds: a systematic map and user guide},
journal = {Ecological Applications},
volume = {29},
number = {6},
pages = {e01954},
url = {https://esajournals.onlinelibrary.wiley.com/doi/abs/10.1002/eap.1954},
year = {2019}
}

@article{Fairbrass2017,
title = {Biases of acoustic indices measuring biodiversity in urban areas},
journal = {Ecological Indicators},
volume = {83},
pages = {169-177},
year = {2017},
doi = {https://doi.org/10.1016/j.ecolind.2017.07.064},
author = {Alison J. Fairbrass and Peter Rennert and Carol Williams and Helena Titheridge and Kate E. Jones}
}

@article{Gibb2019,
author = {Gibb, Rory and Browning, Ella and Glover-Kapfer, Paul and Jones, Kate E.},
title = {Emerging opportunities and challenges for passive acoustics in ecological assessment and monitoring},
journal = {Methods in Ecology and Evolution},
volume = {10},
number = {2},
pages = {169-185},
doi = {https://doi.org/10.1111/2041-210X.13101},
year = {2019}
}

@article{Hoefer2023,
author = {Hoefer, Sebastian and McKnight, Donald and Allen-Ankins, Slade and Nordberg, Eric and Schwarzkopf, Lin},
year = {2023},
pages = {506-531},
title = {Passive acoustic monitoring in terrestrial vertebrates: a review},
volume = {32},
journal = {Bioacoustics},
doi = {10.1080/09524622.2023.2209052}
}

@article{Hingston2018,
author = {Hingston, Andrew and Wardlaw, Tim and Baker, Susan and Jordan, Gregory},
year = {2018},
pages = {30-39},
title = {Data obtained from acoustic recording units and from field observer point counts of Tasmanian forest birds are similar but not the same},
volume = {35},
journal = {Australian Field Ornithology},
doi = {10.20938/afo35030039}
}

@article{Hughes2022,
author = {Hughes, Alice and Orr, Michael and Lei, Fumin and Yang, Qinmin and Qiao, Huijie},
year = {2022},
title = {Understanding drivers of global urban bird diversity},
volume = {76},
journal = {Global Environmental Change},
doi = {10.1016/j.gloenvcha.2022.102588}
}

@article{Robinson2000,
author = {Robinson, W. and Brawn, Jeffrey and Robinson, Scott},
year = {2000},
pages = {209-235},
title = {Forest Bird Community Structure in Central Panama: Influence of Spatial Scale and Biogeography},
volume = {70},
journal = {Ecological Monographs},
doi = {https://doi.org/10.1890/0012-9615(2000)070[0209:FBCSIC]2.0.CO;2},
}

@article{Robinson2018,
author = {Robinson, W. Douglas and Lees, Alexander C. and Blake, John G.},
title = {Surveying tropical birds is much harder than you think: a primer of best practices},
journal = {Biotropica},
volume = {50},
number = {6},
pages = {846-849},
doi = {https://doi.org/10.1111/btp.12608},
year = {2018}
}

@article{Wheeldon2019,
author = {Wheeldon, Amie and Mossman, Hannah L. and Sullivan, Martin J. P. and Mathenge, James and de Kort, Selvino R.},
title = {Comparison of acoustic and traditional point count methods to assess bird diversity and composition in the Aberdare National Park, Kenya},
journal = {African Journal of Ecology},
volume = {57},
number = {2},
pages = {168-176},
doi = {https://doi.org/10.1111/aje.12596},
year = {2019}
}

@article{Wong2023,
author = {Wong, Joy and Soh, Malcolm and Low, Bing Wen and Er, Kenneth},
year = {2023},
title = {Tropical bird communities benefit from regular-shaped and naturalised urban green spaces with water bodies},
volume = {231},
journal = {Landscape and Urban Planning},
doi = {10.1016/j.landurbplan.2022.104644}
}

@article{Gupta2021,
author = {Gupta, Gaurav and Kshirsagar, Meghana and Zhong, Ming and Gholami, Shahrzad and Lavista Ferres, Juan},
year = {2021},
title = {Comparing recurrent convolutional neural networks for large scale bird species classification},
volume = {11},
journal = {Scientific Reports},
doi = {10.1038/s41598-021-96446-w}
}

@article{Briggs2012,
author = {Briggs, Forrest and Lakshminarayanan, Balaji and Neal, Lawrence and Fern, Xiaoli and Raich, Raviv and Frey, Sarah and Hadley, Adam and Betts, Matthew},
year = {2012},
pages = {4640-50},
title = {Acoustic classification of multiple simultaneous bird species: A multi-instance multi-label approach},
volume = {131},
journal = {The Journal of the Acoustical Society of America},
doi = {10.1121/1.4707424}
}

@article{Thakur2019,
    author = {Thakur, Anshul and Thapar, Daksh and Rajan, Padmanabhan and Nigam, Aditya},
    title = "{Deep metric learning for bioacoustic classification: Overcoming training data scarcity using dynamic triplet loss}",
    journal = {The Journal of the Acoustical Society of America},
    volume = {146},
    number = {1},
    pages = {534-547},
    year = {2019},
    month = {07},
    issn = {0001-4966},
    doi = {10.1121/1.5118245},
}

@article{LeBien2020,
title = {A pipeline for identification of bird and frog species in tropical soundscape recordings using a convolutional neural network},
journal = {Ecological Informatics},
volume = {59},
year = {2020},
issn = {1574-9541},
doi = {https://doi.org/10.1016/j.ecoinf.2020.101113},
url = {https://www.sciencedirect.com/science/article/pii/S1574954120300637},
author = {Jack LeBien and Ming Zhong and Marconi Campos-Cerqueira and Julian P. Velev and Rahul Dodhia and Juan Lavista Ferres and T. Mitchell Aide}
}

@article{Manriquez2024,
author = {Rodrigo Manriquez P, Sonja A. Kotz, Andrea Ravignani and Bart de Boer},
title = {Bioacoustic classification of a small dataset of mammalian vocalisations using deep learning},
journal = {Bioacoustics},
volume = {33},
number = {4},
pages = {354--371},
year = {2024},
publisher = {Taylor \& Francis},
doi = {10.1080/09524622.2024.2354468},
URL = {https://doi.org/10.1080/09524622.2024.2354468},
}

@article{Nolasco2023,
title = {Learning to detect an animal sound from five examples},
journal = {Ecological Informatics},
volume = {77},
year = {2023},
issn = {1574-9541},
doi = {https://doi.org/10.1016/j.ecoinf.2023.102258},
author = {Ines Nolasco and Shubhr Singh and Veronica Morfi and Vincent Lostanlen and Ariana Strandburg-Peshkin and Ester Vidaña-Vila and Lisa Gill and Hanna Pamuła and Helen Whitehead and Ivan Kiskin and Frants H. Jensen and Joe Morford and Michael G. Emmerson and Elisabetta Versace and Emily Grout and Haohe Liu and Burooj Ghani and Dan Stowell},
}

@article{Noumida2022,
title = {Multi-label bird species classification from audio recordings using attention framework},
journal = {Applied Acoustics},
volume = {197},
year = {2022},
issn = {0003-682X},
doi = {https://doi.org/10.1016/j.apacoust.2022.108901},
url = {https://www.sciencedirect.com/science/article/pii/S0003682X22002754},
author = {A Noumida and Rajeev Rajan}
}

@article{schwinger2025,
title = {Foundation models for bioacoustics – A comparative review},
journal = {Ecological Informatics},
volume = {96},
pages = {103765},
year = {2026},
issn = {1574-9541},
doi = {https://doi.org/10.1016/j.ecoinf.2026.103765},
url = {https://www.sciencedirect.com/science/article/pii/S1574954126001718},
author = {Raphael Schwinger and Paria {Vali Zadeh} and Lukas Rauch and Mats Kurz and Tom Hauschild and Sam Lapp and Sven Tomforde},
}

@article{Ghani_2023,
   title={Global birdsong embeddings enable superior transfer learning for bioacoustic classification},
   volume={13},
   ISSN={2045-2322},
   url={http://dx.doi.org/10.1038/s41598-023-49989-z},
   DOI={10.1038/s41598-023-49989-z},
   number={1},
   journal={Scientific Reports},
   publisher={Springer Science and Business Media LLC},
   author={Ghani, Burooj and Denton, Tom and Kahl, Stefan and Klinck, Holger},
   year={2023},
   month=dec }

@inproceedings{BirdSet,
  title     = {BirdSet: A Large-Scale Dataset for Audio Classification in Avian Bioacoustics},
  author    = {Lukas Rauch and Raphael Schwinger and Moritz Wirth and René Heinrich and Denis Huseljic and Marek Herde and Jonas Lange and Stefan Kahl and Bernhard Sick and Sven Tomforde and Christoph Scholz},
  booktitle = {International Conference on Learning Representations (ICLR)},
  year      = {2025},
  url       = {https://openreview.net/forum?id=dRXxFEY8ZE}
}

@article{Martínez2025,
author = {Martínez-Medina, Daniela},
year = {2025},
month = {06},
pages = {1-4},
title = {Tuning into nature: the sonic boost transforming tropical biodiversity research},
volume = {380},
journal = {Philosophical Transactions B},
doi = {10.1098/rstb.2024.0044}
}

@article{Brandes2008,
  author={Brandes, T. Scott},
  journal={IEEE Transactions on Audio, Speech, and Language Processing}, 
  title={Feature Vector Selection and Use With Hidden Markov Models to Identify Frequency-Modulated Bioacoustic Signals Amidst Noise}, 
  year={2008},
  volume={16},
  number={6},
  pages={1173-1180},
  doi={10.1109/TASL.2008.925872}}

@misc{birdnet_analyzer_git,
  author       = {BirdNET Team},
  title        = {BirdNET-Analyzer: Neural network for bird sound recognition},
  howpublished = {\url{https://github.com/birdnet-team/BirdNET-Analyzer}},
  note         = {Version 2.4, accessed 8 September 2025}
}

@misc{nparksBirdsSingapore,
  author       = {{National Parks Board}},
  title        = {Checklist of Bird Species for Singapore},
  howpublished = {\url{https://www.nparks.gov.sg/nature/species-list/birds}},
  year         = {2026},
  note         = {Accessed: 2026-06-06}
}

@misc{hexeberg2026,
      title={Time-frequency localization of bird calls in dense soundscapes}, 
      author={Simen Hexeberg and Fanghui Tong and Hari Vishnu and Mandar Chitre},
      year={2026},
      eprint={2606.10407},
      archivePrefix={arXiv},
      primaryClass={cs.SD},
      url={https://arxiv.org/abs/2606.10407}, 
}

@misc{xenocanto,
  title        = {Xeno-canto: An Online Database of Bird Sounds},
  howpublished = {\url{https://xeno-canto.org}},
  note         = {Accessed: 2023}
}

@article{both2012,
  author  = {Both, Christiaan and Grant, Tom},
  title   = {Biological invasions and the acoustic niche: the effect of bullfrog calls on the acoustic signals of white-banded tree frogs},
  journal = {Biology Letters},
  volume  = {8},
  number  = {5},
  pages   = {714--716},
  year    = {2012},
  doi     = {10.1098/rsbl.2012.0412}
}

@article{medeiros2017,
  author  = {Medeiros, Camila Ineu and Both, Camila and Grant, Taran and Hartz, Sandra Maria},
  title   = {Invasion of the acoustic niche: variable responses by native species to invasive American bullfrog calls},
  journal = {Biological Invasions},
  year    = {2017},
  volume  = {19},
  number  = {2},
  pages   = {675--690},
  doi     = {10.1007/s10530-016-1327-7},
  url     = {https://doi.org/10.1007/s10530-016-1327-7},
  issn    = {1573-1464}
}

@article{hopkins2023,
  author  = {Hopkins, J. M. and Bower, D. S. and Edwards, W. and Schwarzkopf, L.},
  title   = {Effects of invasive toad calls and synthetic tones on call properties of native Australian toadlets},
  journal = {Journal of Herpetology},
  volume  = {57},
  pages   = {437--446},
  year    = {2023},
  doi     = {10.1670/23-004}
}

@article{farina2012,
author = {Almo Farina and Nadia Pieretti and Niki Morganti},
title = {Acoustic patterns of an invasive species: the Red-billed Leiothrix (Leiothrix lutea Scopoli 1786) in a Mediterranean shrubland},
journal = {Bioacoustics},
volume = {22},
number = {3},
pages = {175--194},
year = {2013},
publisher = {Taylor \& Francis},
doi = {10.1080/09524622.2012.761571},
URL = {https://doi.org/10.1080/09524622.2012.761571},
eprint = {https://doi.org/10.1080/09524622.2012.761571}
}

@article {Bermant2022,
	author = {Bermant, Peter C. and Brickson, Leandra and Titus, Alexander J.},
	title = {Bioacoustic Event Detection with Self-Supervised Contrastive Learning},
	elocation-id = {2022.10.12.511740},
	year = {2022},
	doi = {10.1101/2022.10.12.511740},
	publisher = {Cold Spring Harbor Laboratory},
	URL = {https://www.biorxiv.org/content/early/2022/10/16/2022.10.12.511740.1},
	eprint = {https://www.biorxiv.org/content/early/2022/10/16/2022.10.12.511740.1.full.pdf},
	journal = {bioRxiv}
}

@INPROCEEDINGS{Moummad2024,
  author={Moummad, Ilyass and Farrugia, Nicolas and Serizel, Romain},
  booktitle={2024 IEEE International Conference on Acoustics, Speech, and Signal Processing Workshops (ICASSPW)}, 
  title={Self-Supervised Learning for Few-Shot Bird Sound Classification}, 
  year={2024},
  volume={},
  number={},
  pages={600-604},
  doi={10.1109/ICASSPW62465.2024.10627576}}

@INPROCEEDINGS{neal2011,
  author={Neal, Lawrence and Briggs, Forrest and Raich, Raviv and Fern, Xiaoli Z.},
  booktitle={2011 IEEE International Conference on Acoustics, Speech and Signal Processing (ICASSP)}, 
  title={Time-frequency segmentation of bird song in noisy acoustic environments}, 
  year={2011},
  volume={},
  number={},
  pages={2012-2015},
  doi={10.1109/ICASSP.2011.5946906}}

@inproceedings{ruiz2015,
author = {Ruiz-Mu\~{n}oz, J. F. and Orozco-Alzate, Mauricio and Castellanos-Dominguez, G.},
title = {Multiple instance learning-based birdsong classification using unsupervised recording segmentation},
year = {2015},
isbn = {9781577357384},
publisher = {AAAI Press}, 
booktitle = {Proceedings of the 24th International Conference on Artificial Intelligence},
pages = {2632–2638},
numpages = {7},
location = {Buenos Aires, Argentina},
series = {IJCAI'15}
}

\end{document}